\documentclass[twocolumn,
	aps, prd,
	10pt, notitlepage, 
        floats, floatfix,
	amsmath, amssymb, amsfonts, eqsecnum,
	superscriptaddress,
	showpacs, showkeys,
	nofootinbib,
 	longbibliography,
]{revtex4-2}

\usepackage{mathtools}
\usepackage{wrapfig}
\usepackage{graphicx} 
\usepackage[usenames,dvipsnames]{xcolor}
\usepackage{xspace} 
\usepackage{bm} 
\usepackage{amsmath,amsfonts,amssymb,amsthm,mathtools}
\usepackage[utf8]{inputenc} 
\usepackage{subcaption}

\xdefinecolor{mylinkcolor}{rgb}{0,0,0.5}
\usepackage[
	bookmarksnumbered, bookmarksopen, bookmarksopenlevel=2,
	breaklinks=true,
	colorlinks=true, filecolor=mylinkcolor, citecolor=mylinkcolor,
	linkcolor=mylinkcolor, urlcolor=mylinkcolor, menucolor=mylinkcolor,
]{hyperref}

\usepackage{verbatim}
\usepackage{mathrsfs}
\usepackage{color}
\usepackage{enumitem}
\usepackage[dvipsnames]{xcolor}
\usepackage{comment}
\usepackage{tikz}
\usetikzlibrary{decorations.pathmorphing}
\usetikzlibrary{arrows}

\tikzset{snake it/.style={decorate, decoration=snake}}

\usepackage[normalem]{ulem}

\newcommand{\nnm}{\nonumber}

\newcommand{\be}{\begin{equation}}
\newcommand{\ee}{\end{equation}}
\newcommand{\bse}{\begin{subequations}}
\newcommand{\ese}{\end{subequations}}

\newcommand{\mr}{\mathrm}

\newcommand{\mce}{\mathcal{E}}
\newcommand{\mcb}{\mathcal{B}}
\newcommand{\bpm}{\begin{pmatrix}}
\newcommand{\epm}{\end{pmatrix}}
\newcommand{\hs}{\hat{s}}
\newcommand{\hS}{\hat{S}}
\newcommand{\AEI}{\affiliation{Max Planck Institute for Gravitational Physics (Albert Einstein Institute), Am M\"uhlenberg 1, Potsdam 14476, Germany}}
\newcommand{\Maryland}{\affiliation{Department of Physics, University of Maryland, College Park, MD 20742, USA}}

\begin{document}
\title{Modeling horizon absorption in spinning binary black holes\\using effective worldline theory}
\author{M. V. S. Saketh}
\email{msaketh@umd.edu}
\Maryland
\AEI
\author{Jan Steinhoff}
\email{jan.steinhoff@aei.mpg.de}
\AEI
\author{Justin Vines}
\email{justin.vines@aei.mpg.de}
\AEI
\author{Alessandra Buonanno}
\email{alessandra.buonanno@aei.mpg.de}
\AEI
\Maryland
\begin{abstract}

 The mass and spin of black holes (BHs) in binary systems
  may change due to the infall of gravitational-wave (GW) energy down the
  horizons. For spinning BHs, this effect enters at 2.5
  post-Newtonian (PN) order relative to the leading-order energy flux
  at infinity. There is currently a discrepancy in the literature in
  the expressions of these horizon fluxes in the test-body limit at
  4PN order (relative 1.5PN order). Here, we model the horizon absorption as
  tidal heating in an effective worldline theory of a spinning
  particle equipped with tidally-induced quadrupole and octupole
  moments. We match the tidal response to analytic solutions of the
  Teukolsky equation in a scattering scenario, and obtain general
  formulae for the evolution of mass and spin. We then 
  specialize to the case of aligned-spin--quasi-circular 
  binaries, obtaining the corresponding contributions to
  the GW phasing through 4PN order. Importantly, we find that 
the number of GW cycles due to horizon fluxes with masses observed by LIGO-Virgo-KAGRA detectors is about 2-3 orders of magnitude smaller than the other contributions to the phasing at the same PN order.
Furthermore, in the test-body
  limit, we find full agreement with results obtained earlier from
  BH perturbation theory, with a small mass in an equatorial
  circular orbit treated as a source perturbing the Kerr metric. Thus,  
we weigh in on one side of the previous discrepancy.
\end{abstract}
\maketitle

\section{Introduction and summary}
With ninety gravitational-wave (GW) events~\cite{LIGOScientific:2021djp} from compact-binary coalescences observed by the LIGO-Virgo detectors~\cite{LIGOScientific:2014pky,VIRGO:2014yos}, GW astronomy has become an important instrument to explore our universe.
While the worldwide network of detectors has recently included the KAGRA detector~\cite{KAGRA:2020agh}, and will continue to improve in sensitivity in the future~\cite{Saleem:2021iwi,LIGOScientific:2016wof,Punturo:2010zza,LISA:2017pwj,Reitze:2019iox}, the accuracy of waveform models needs to keep in step with the increasing sensitivity in order to avoid systematic errors in parameter estimation~\cite{Purrer:2019jcp}.
Predictions for GWs from the late stage of a binary inspiral and merger require full numerical solutions of the strong-field dynamics.
Here we focus instead on the early inspiral, applying the post-Newtonian (PN, weak-field and slow motion) approximation, and calculate horizon-absorption effects on the GW phase from black hole (BH) binaries up to 4PN order.

The presence of the horizon leads to some interesting effects in BH binaries. In particular, GW energy can fall into (or out of, in the case of superradiance) the horizon, leading to a change in mass and magnitude of spin angular momentum\footnote{We will refer to the magnitude of spin angular momentum simply as ``spin" in the rest of this work.} of the BHs in a binary. The change in these parameters is consistent with the second law of BH dynamics and always leads to an increase (or no change) in the area of the horizon~\cite{Hawking:1971vc}. A change in mass and spin can also happen for other compact bodies like neutron stars through tidal heating (see, e.g., Refs.~\cite{Meszaros:1992ps,Lai:1993di}). Future GW detectors may be sensitive to these dissipative effects, which can then provide a probe for the nature of BHs (e.g., the presence of a horizon)~\cite{Maselli:2017cmm,Datta:2019epe, Datta:2019euh, Mukherjee:2022wws}. Horizon-absorption effects have been included in some effective-one-body (EOB) waveform models~\cite{Nagar:2011aa,Bernuzzi:2012ku,Taracchini:2013wfa}, but not yet in the state-of-the-art models used for LIGO-Virgo-KAGRA (LVK) data analysis.

For spinning BHs, the leading-order flux into the horizon starts at 2.5PN with respect to the leading quadrupolar flux (of the binary) to infinity~\cite{Tagoshi:1997jy, Chatziioannou:2016kem, Goldberger:2020fot,Poisson:2005pi, Comeau:2009bz, Poisson:2004cw, Yunes:2005ve}. Through the flux-phase relation, this means that the GW phase is also affected at 2.5PN order (see, e.g., Ref.~\cite{Brown:2007jx,Isoyama:2017tbp}). For nonspinning BHs, the same effect starts at 4PN~\cite{Tagoshi:1997jy, Poisson:2005pi}. In the test-body limit, when there is a tiny BH orbiting a much larger spinning BH, one can solve for the horizon energy flux of the large BH via BH perturbation-theory (BHPT), which consists in this case of solving the Teukolsky equation with incoming boundary conditions at the horizon, and outgoing boundary conditions at infinity, as was done, e.g., in Ref.~\cite{Tagoshi:1997jy}. However, it is nontrivial to extend this calculation to generic mass ratios. This was accomplished at leading 2.5PN order in Ref.~\cite{Alvi:2001mx}, for the case of aligned-spin circular orbits via arguments relating the spin-aligned--quasi-circular inspiral to the case when two spinning BHs are held at rest with respect to each other. The result was extended further to 1.5PN orders (to absolute 4PN order) in Refs.~\cite{Chatziioannou:2012gq, Chatziioannou:2016kem} where BHPT was used along with a matching between the near zone and the orbital zone for BHs in a binary to derive the energy and angular-momentum fluxes across their horizons for generic mass ratios. However, for spinning BHs, the result was inconsistent with that obtained in the test-body limit in Ref.~\cite{Tagoshi:1997jy}. This discrepancy is yet to be settled in the literature and thus the correct expression for mass and angular-momentum evolution for generic mass ratios at 4PN is yet to be clarified. Settling this discrepancy  is crucial to derive the correct horizon-flux contribution to the waveform phase at 4PN order,  and it is one of the goals of this paper.

The problem of computing the horizon fluxes for spinning BHs was tackled recently in an effective field theory (EFT) framework in Ref.~\cite{Goldberger:2020fot} (see also, e.g., Refs.~\cite{Goldberger:2005cd,Porto:2007qi,Goldberger:2012kf,Endlich:2015mke,Endlich:2016jgc,Goldberger:2020wbx} for previous work in that direction and Refs.~\cite{Goldberger:2004jt,Porto:2005ac,Goldberger:2009qd,Levi:2015msa,Delacretaz:2014oxa,Gupta:2020lnv} for the EFT formalism including spin and tidal effects), where the BH was treated as a point particle with tidally induced quadrupole moments. Then in an ``in-in" formalism, a parametrized expression for the absorption cross section for a graviton being absorbed by the effective particle was computed, whose parameters were fixed by a matching calculation against the classical absorption cross section for gravitational plane waves by a spinning BH(from Ref.~\cite{Page:1976df}). This in turn fixes the correlation function for the quadrupole moment, which was subsequently used to derive the dissipative part of the Green's function relating the tidal fields to the quadrupole moments. Once the tidal response was fixed in this way, the EFT was applied to compute the evolution equation for mass and spin by suitably defining them in the worldline theory from the effective action. This calculation was carried out at leading (2.5PN) order and the result was consistent with earlier works in Refs.~\cite{Poisson:2005pi, Comeau:2009bz, Poisson:2004cw, Yunes:2005ve}. 

In this work, we similarly develop an effective worldline framework to model the dissipative dynamics of spinning BHs, but extend it by 1.5PN orders. We follow Ref.~\cite{Goldberger:2020fot} in treating the spinning BH as a point particle with tidally induced moments, but work in a purely classical framework. We identify and fix the dissipative part of the tidal response by comparing results from the scattering of GWs off the particle/BH between the effective and real theories (see also, e.g., Refs.~\cite{Bautista:2021wfy,Creci:2021rkz,Saketh:2022wap,Ivanov:2022qqt} for other works involving comparison of such a scattering process between real and effective theories).
We then derive expressions for the evolution of mass and spin up to 1.5PN order (relative to the leading order). We find that the resulting expressions are consistent with earlier results obtained in the test-body limit in Ref.~\cite{Tagoshi:1997jy}, thus weighing in on one side of the discrepancy discussed above. We then derive the 4PN contribution to the waveform phase due to the horizon fluxes, while consistently including the effects due to the changing parameters (mass and spin) of the members of the binary (extending an earlier 3.5PN result~\cite{Isoyama:2017tbp}).

To accomplish this, we write down  in Sec.~\ref{sec:EFT} an effective worldline action for a spinning particle with (gravito-electric and -magnetic) quadrupole and octupole moments, coupled accordingly to quadrupolar and octupolar (gravito-electric and -magnetic) tidal fields in the action. We then motivate ans{\"a}tze relating the tidal fields linearly to the multipole moments. In the absence of spin,  spherical symmetry and parity symmetry imply that a given tidal field only induces the corresponding multipole moment (e.g., the electric-type quadrupole $Q_\mathcal{E}^{\mu\nu}$ is induced only by the electric-type tidal field $\mathcal{E}^{\mu\nu}$). However, in the presence of spin, it is possible for octupolar tidal fields to induce quadrupolar tidal fields and vice-versa, while still preserving parity. This is an important property of the ans{\"a}tze and turns out to be crucial for correctly modelling the dissipative dynamics of spinning BHs. In particular, the Teukolsky equation which governs the curvature perturbations in a Kerr background is separable in spheroidal harmonics with spin weight -2. This feature can be modelled in the effective theory only by including the interaction between quadrupole (octupole) fields and octupole (quadrupole) moments.

Once we motivate general ans{\"a}tze for the multipole moments, we further specialize them by using the fact that the response tensors (relating the tidal fields to the multipole moments) can only have a nontrivial tensor structure due to the spin of the particle, which allows us to decompose them into a set of basis tensors with undetermined coefficients to be fixed. For this purpose, in Sec.~\ref{sec:scatter}, we place the effective particle at the origin and scatter GWs off of it, and then use the Einstein equation and the ans{\"a}tze to solve for the scattered wave and then subsequently to compute the degree of absorption for the spheroidal $l=2,~3$ modes of the wave to $\mathcal{O}(\epsilon^7)$, where $\epsilon=GM\omega$ with $M$ being the BH mass and $\omega$ the GW angular frequency. Comparing the degree of absorption obtained by solving the scattering problem in the effective theory with that obtained by solving the same problem in the actual setup of GWs scattering off a spinning BH, using BHPT as governed by the Teukolsky equation \cite{Teukolsky:1972my}, finally fixes the response coefficients that contribute to dissipation.  The response coefficients are notably nonpolynomial in the BH spin.

Once the (dissipative part of the) tidal response is fixed, we proceed to compute expressions for the evolution of mass and spin in the effective theory in Sec.~\ref{dmjdt}. We first derive general evolution equations for mass $m$, and spin $J$, in terms of the tidal fields and multipole moments from the equations of motion obtained from the action, and then derive the explicit expressions for the special case parallel-spin--quasi-circular\footnote{In this work, by parallel-spin--quasi-circular binaries, we always mean BHs in a binary with their spin vectors parallel to each other and to the orbital angular momentum.} binaries to relative 1.5PN order. We show its consistency (or lack thereof) with earlier results and then proceed to compute the effect on the waveform phase up to 4PN with respect to the leading-order quadrupolar flux of the system to infinity in Sec.~\ref{sec:phase}. With that, we conclude our work in this paper in Sec.~\ref{sec:conclude}.

We work with the (-,+,+,+) metric signature convention. We use greek symbols $\mu$, $\nu$, ..., for space-time indices ranging over $\{0,1,2,3\}$ with 0 being used for the time-component, and latin symbols $i$, $j$, ..., for spatial indices ranging over $\{1,2,3\}$. We use $\epsilon_{0123}=\epsilon_{123}=1$ as the convention for Levi-Civita tensor(s). We also use the multi-index notation where $\mu_L= \mu_1\mu_2\dots \mu_l$ for conveniently representing a string of indices where useful and use the notation $\langle \mu_1\mu_2...\rangle$ to represent symmetrization and trace removal of a tensor over the contained indices. We set the speed of light $c=1$ in the work. However, we keep the dependence on the Newton constant $G$ explicit for most of the work, and mention explicitly when setting $G=1$ as well to facilitate comparison with earlier works.
\section{Setup in effective worldline theory}\label{sec:EFT}
In this section, we discuss the effective worldline theory for the point particle used to model the spinning compact object. We first briefly outline the particle's multipole structure and how it can be used to model absorption. We then proceed to set up an effective action for the particle including the multipole moments, and then write down general ans{\"a}tze with undetermined coefficients for the multipole moments as a linear function of the tidal fields consistent with axisymmetry and parity symmetry. 

We model the horizon flux of the spinning BH in effective worldline theory as tidal heating of a composite paricle with several tidally induced multipole moments, whose degrees of freedom are contained in symmetric trace-free (STF) tensors $Q^{\mu_L}_{\mce,n}$, $Q^{\mu_L}_{\mcb,n}$, $l\geq 2$, $n\geq 0$, satisfying $Q^{\mu_L}u_{\mu_i}=0$, $1\leq i \leq l$. In the effective action, we choose the multipole moments to couple with the tidal fields
\begin{alignat}{3}
\label{tflds}
\mce_{\mu_L} &= \nabla_{\langle\mu_{L-2}}R_{\mu_{l-1}|\alpha|\mu_l\rangle\beta}u^{\alpha}u^{\beta},  \\ \nnm \mcb_{\mu_L} &= \frac{1}{2}\nabla_{\langle\mu_{L-2}}\epsilon_{|\gamma|\mu_{l-1}}{}^{\alpha\beta}R_{|\alpha\beta|\mu_l\rangle\delta}u^{\gamma}u^{\delta}, ~l\geq 2,
\end{alignat}
 where $u^{\mu}$ is the four-velocity and we are using the notation $\langle\mu_L\rangle$ to denote symmetrization and trace-removal. In the effective action, we choose the electric `$\mce$' (magnetic `$\mcb$') multipole moments to couple with electric (magnetic) tidal fields in accordance with the number of indices and parity as  $S_{\mr{tidal}} = (1/2)\sum_{l=2}^{\infty}\sum_n [Q_{\mce,n}^{\mu_L}\mce_{\mu_L} + (\mce\leftrightarrow \mcb)]]$. We will only need to consider their coupling and induction by quadrupolar ($l=2$) and octupolar ($l=3$) tidal fields to the PN order relevant in this work. In addition, the various multipole moments are dynamical and are coupled to each other via an internal action $S_{\mr{int}}$ such that the energy tidally pumped into these modes may progressively escape into higher order/smaller length-scale multipoles effectively leading to dissipation. If there is a sufficiently large number of degrees of freedom, and if they are appropriately coupled, the recurrence time becomes essentially infinite and the system becomes effectively irreversible. We will not however explicitly model the process of dissipation and only use that as a justification to write down ans{\"a}tze for the multipole moments in terms of the tidal fields that allows for dissipation. Note that at the end of the day, we only intend to mimic the BH's horizon absorption (as tidal heating) and acquire an effective model that may be used to study the associated dynamics. Whether there is any physical relation to the real microscopic degrees of freedom of a BH and this model is unknown and not directly relevant to this work. Our approach of incorporating tidal moments in the action is slightly different at a superficial level from the prescription used in Ref.~\cite{Goldberger:2020fot} where instead a single quadrupole moment was used but allowed to be a function of several unknown microscopic degrees of freedom denoted by $X$. Practically however, there is not much of a difference.

 In the absence of spin, an unperturbed BH is spherically symmetric, and the linear tides can only be induced by the fields to which they directly couple to in the action, e.g., as\footnote{We will only ever need an ansatz for the sum of all multipole moments for a given $l$ and parity ($\mce/\mcb$) for computing the evolution equations for physical quantities such as total spin angular momentum or total linear momentum. We however allow for the presence of multiple multipole moments with the same $l$ and parity labels $\mce/\mcb$ for generality.},
\begin{alignat}{3}
\label{scwmom}
 & \sum_{n=0}^{\infty}  Q^{\mu_L}_{\mce,n} = M\sum_{m=0}^{\infty}\lambda_{\mce,m}^l(GM)^{2+l+m}\frac{D^m}{D\tau^m}\mce^{\mu_L},  \nnm \\ & \text{and similarly for } (\mce\leftrightarrow \mcb)
\end{alignat}
where parity symmetry\footnote{In general, $\mce_{\mu_L}$ (or $\mcb_{\mu_L}$) transforms under parity as $(-1)^{l}$ or $[(-1)^{(l+1)}]$. The multipole moments with which they explicitly couple in the action cannot be induced by tidal fields with a different transformation under parity if the particle is to be parity-preserving.} prevents the magnetic (electric) tidal field for the same $l$ from inducing the electric (magnetic) multipole moment, and spherical symmetry means there is no special tensor with which to contract the higher multipolar order tidal fields (or multiply the lower multipolar order ones) to contribute to the ansatz for $\sum_n Q^{\mu_L}_{\mce(\mcb),n}$. Spherical symmetry is also the reason there is no mixing of indices in the response tensor. We can also identify from Eq.~(\ref{scwmom}) which response coefficients are conservative and which are dissipative simply by looking at the transformation under time-reversal. The coefficients next to odd powers of time derivatives are dissipative and the ones next to even powers are conservative. This is less trivial for a spinning particle which allows for mixing of multipolar orders with the help of the spin tensor and Pauli-Lubanski spin vector.

However, a spinning BH can have induction between different multipolar orders (different $l$), since the spin-tensor $S^{\mu\nu}$ defined in Eq.~(\ref{sap}), or Pauli-Lubanski spin vector $s^{\mu} =- [1/(2m)]\epsilon^{\mu}{}_{\nu\rho\sigma}p^{\nu}S^{\rho\sigma}$, can appear in above relations~\eqref{scwmom}. We still need to keep parity considerations in mind as spinning BHs obey parity symmetry. Crucially, for our purposes, we need to include the induction of octupole moments by quadrupolar fields and vice-versa to capture the tidal heating of the spinning BHs at relative 1.5PN order. We will write down a general ansatz for multipole moments in the spinning case in Subsec.~\ref{nyayam}. But first, we will write down in Subsec.~\ref{action_principle} an effective action for a particle with spin and aforementioned multipole moments and derive the equations of motion for spin and four-momentum, and the effective stress-energy tensor, following closely the prescription in Ref.~\cite{Marsat:2014xea} but with suitable modifications to allow for the presence of tidally induced moments. We will then write down general parametrized ans{\"a}tze for the multipole moments in terms of the tidal fields constrained by the symmetries of the particle (axisymmetry and parity invariance), which is somewhat similar to the approach used in Ref.~\cite{Goldberger:2020fot} to fix the form of the correlation function of the quadrupole moments. 
\subsection{Action and equations of motion for momentum and spin angular momentum}
\label{action_principle}
To get the equations of motion, we will follow a direct extension of the simple procedure given in Ref.~\cite{Marsat:2014xea} for deriving the equations of motion from an implicit action, while including the aforementioned tidal moments. We include the spinning degrees of freedom by attaching to the particle a body-fixed tetrad $\epsilon_A{}^{\mu}$ satisfying orthonormality and completeness. The angular velocity of the particle is then measured with the quantity $\Omega^{\mu\nu} = \epsilon_A{}^{\mu} \frac{D \epsilon^{A\nu}}{D\tau}$. It is sufficient in this work just to include the spin at leading order and ignore spin-induced multipole moments. The worldline of the particle is denoted by $z^{\mu}(\tau)$, where $\tau$ is the proper time.
We can then write down the action implicitly as 
\begin{alignat}{3}
\nnm S = \int d\tau L(& u^{\mu},\Omega^{\mu\nu},g_{\mu\nu}, Q^{\mu_L}_{\mce,n},Q^{\mu_L}_{\mcb,n},\dot{Q}^{\mu_L}_{\mce,n}, \dot{Q}^{\mu_L}_{\mcb,n},\\& R_{\mu\nu\rho\sigma},\nabla_{\lambda}R_{\mu\nu\rho\sigma}),
\end{alignat}
where $u^{\mu}=dz^{\mu}/d\tau$ and we are using the notation $\dot{a} = Da/D\tau$. Additionally, we have assumed that only the first time derivative of each individual multipole moment needs to be included in the action. We have also restricted ourselves to including just the quadrupolar ($l=2$) and octupolar ($l=3$) tidal fields in the action explicitly as mentioned before. The coupling of tidal fields with higher order multipole moments is irrelevant to the PN order of interest in this work as we will see later. We do however keep all the higher multipolar order moments since tidal heating requires the presence of several additional degrees of freedom into which the system may pump energy. Additionally, as mentioned before, we choose the multipole moments to couple directly with the corresponding tidal fields (in accordance with number of indices and parity label) by imposing
\begin{alignat}{3}
\label{dcoup} & \frac{\partial L}{\partial \mce_{\mu_L}} = \frac{1}{2}\sum_n Q_{\mce,n}^{\mu_L},~\frac{\partial L}{\partial \mcb_{\mu_L}} = \frac{1}{2}\sum_n Q_{\mcb,n}^{\mu_L}, \quad  l=2,~3 \nnm \\&
\end{alignat}
which is equivalent to having in the action a linear combination of the form $S_{\mr{tidal}} =\sum_{l=2}^{3}\sum_n \int d\tau \frac{1}{2}\mce_{\mu_L}Q_{\mce,n}^{\mu_L} + (\mce\leftrightarrow \mcb)$ in the total action. We can derive the equations of motion for momentum and spin angular momentum directly from this implicit action. First, we consider a general variation 
\begin{widetext}
\begin{alignat}{3}
\delta L = & p_{\mu}\delta u^{\mu} + \frac{1}{2}S_{\mu\nu}^{\mr{rot}}\delta \Omega^{\mu\nu} + \frac{\partial L}{\partial g_{\mu\nu}}\delta g_{\mu\nu} - \frac{1}{6}J^{\mu\nu\rho\sigma}\delta R_{\mu\nu\rho\sigma} - \frac{1}{12}J^{\lambda\mu\nu\rho\sigma}\delta\nabla_{\lambda}R_{\mu\nu\rho\sigma} \nnm \\& + \sum_{l=2}^\infty\sum_{n=0}^{\infty}\bigg(\frac{\partial L}{\partial Q_{\mce,n}^{\mu_L}}\delta Q_{\mce,n}^{\mu_L}  + \frac{\partial L}{\partial \dot{Q}_{\mce,n}^{\mu_L}}\delta \dot{Q}_{\mce,n}^{\mu_L} +  \frac{\partial L}{\partial Q_{\mcb,n}^{\mu_L}}\delta Q_{\mcb,n}^{\mu_L}  + \frac{\partial L}{\partial \dot{Q}_{\mcb,n}^{\mu_L}}\delta \dot{Q}_{\mcb,n}^{\mu_L}\bigg) ,
\end{alignat}
where we have defined
\begin{alignat}{3}
p_{\mu} &=(\partial L/\partial u^{\mu})|_{\Omega^{\mu\nu}}, ~S^{\mr{rot}}_{\mu\nu} = \frac{1}{2}(\partial L/\partial \Omega^{\mu\nu}), \label{sap} \\& J^{\mu\nu\rho\sigma} =  -6\frac{\partial L}{\partial R_{\mu\nu\rho\sigma}} =- 3 \sum_{n=0}^\infty u^{[\mu}Q_{\mce,n}^{\nu][\rho}u^{\sigma]}+\frac{3}{2}\sum_{n=0}^\infty Q_{\mcb,n}^{\alpha\langle\rho}\epsilon_{\beta\alpha}{}^{\mu\nu}u^{|\beta|}u^{\sigma\rangle_R},\label{4quad}\\& J^{\lambda\mu\nu\rho\sigma} = -12\frac{\partial L}{\partial \nabla_{\lambda}R_{\mu\nu\rho\sigma}} = - 6 \sum_{n=0}^\infty u^{\langle\mu}Q_{\mce,n}^{\nu\rho\lambda}u^{\sigma\rangle_{\nabla R}}+3\sum_{n=0}^\infty Q_{\mcb,n}^{\alpha\langle\rho\lambda}\epsilon_{\beta\alpha}{}^{\mu\nu}u^{|\beta|}u^{\sigma\rangle_{\nabla R}}, \label{4oct}
\end{alignat}
\end{widetext}
where the expressions for the $J$'s follow trivially from Eq.~(\ref{dcoup}) and we are using $\langle abcd \rangle_R$ ($\langle abcd \rangle_{\nabla R}$)  to represent the symmetrization of indices according to the symmetries of the Riemann tensor (covariant derivative of the Riemann tensor) respectively. Now, if we consider a variation of the form $x^{\mu}\rightarrow x^{\mu}+\xi^{\mu}$ corresponding to an infinitesimal change of coordinates, we obtain the constraint 
\begin{widetext}
\begin{alignat}{3}
2 \frac{\partial L}{\partial g_{\mu\nu}} &= p^{\mu}u^{\nu} + S_{\mr{rot}}^{\mu\rho}\Omega^{\nu}{}_{\rho}+ \frac{2}{3}R^{\mu}{}_{\lambda\rho\sigma}J^{\nu\lambda\rho\sigma}+\frac{1}{3}J^{\lambda\nu\tau\rho\sigma}\nabla_{\lambda}R^{\mu}{}_{\tau\rho\sigma}+\frac{1}{12}J^{\nu\lambda\tau\rho\sigma}\nabla^{\mu}R_{\lambda\tau\rho\sigma} \nnm \\& +\sum_{l=2}^{\infty}\frac{l}{2}\sum_{n=0}^{\infty}[\mathcal{M}_{\mce,n}^{\mu~\mu_{L-1}}Q^{\mce,n}_{\mu_{L-1}}{}^{\nu} + P_{\mce,n}^{\mu~\mu_{L-1}}\dot{Q}^{\mce,n}_{\mu_{L-1}}{}^{\nu} + (\mce\leftrightarrow \mcb)],
\label{scon}
\end{alignat}
\end{widetext}
where we have defined $\mathcal{M}_{\mce(\mcb),n}^{\mu_L} = (\partial L/\partial \dot{Q}_{\mu_L}^{\mce(\mcb),n})$, and $P_{\mce(\mcb),n}^{\mu_L}= (\partial L/ \partial Q_{\mu_L}^{\mce(\mcb),n})$. Note that the equations of motion for the multipole moments obtained upon variation of the action with respect to them  $(\partial L/\partial Q^{\mu_L}_{\mce(\mcb),n})- (D/D\tau)(\partial L/\partial \dot{Q}^{\mu_L}_{\mce(\mcb),n})=0$, imply that $\mathcal{M}_{\mce(\mcb),n}^{\mu_L} = \dot{P}^{\mu_L}_{\mce(\mcb),n}$ on the actual worldline.
Eq.~(\ref{scon}) is a useful identity to eliminate the partial derivative with respect to the metric later in the equations of motion for momentum and spin angular momentum. First, we obtain the equation for spin angular momentum easily by variation of the action with respect to the tetrad variables $\epsilon_A{}^{\mu}$. We get the simple equation 
\begin{alignat}{3}
\delta S = \int d\tau \frac{\partial L}{\partial \Omega^{\mu\nu}}\delta\Omega^{\mu\nu}  = 0 \implies \frac{D S^{\mu\nu}_{\mr{rot}}}{D\tau} = 2\Omega^{[\mu}{}_{\rho}S_{\mr{rot}}^{\nu]\rho}, \nnm \\&
\end{alignat}
and we can eliminate the RHS using Eq.~(\ref{scon}) by taking its antisymmetric part [which leads to ($\partial L/\partial g_{\mu\nu}$) vanishing] and we get
\begin{widetext}
\begin{alignat}{3}
\frac{DS^{\mu\nu}_{\mr{rot}}}{D\tau}& = 2 p^{[\mu}u^{\nu]}+\frac{4}{3}R^{[\mu}{}_{\lambda\rho\sigma}J_{\lambda}{}^{\nu]\tau\rho\sigma}+\frac{2}{3}\nabla^{\lambda}R^{[\mu}{}_{\tau\rho\sigma}J_{\lambda}{}^{\nu]\tau\rho\sigma}+\frac{1}{6}\nabla^{[\mu}R_{\lambda\tau\rho\sigma}J^{\nu]\lambda\tau\rho\sigma} \nnm \\& + \sum_{l=2}^{\infty}\sum_{n=0}^{\infty} l\frac{D}{D\tau}[P_{\mce,n}^{[\mu}{}_{\mu_{L-1}}Q^{\nu]\mu_{L-1}}_{\mce,n}+(\mce\leftrightarrow \mcb)],
\end{alignat}
\end{widetext}
where we have used $\mathcal{M}_{\mce(\mcb),n}^{\mu_L} = \dot{P}^{\mu_L}_{\mce(\mcb),n}$, valid on the worldline. We can now redefine the spin angular momentum as $S^{\mu\nu}= S_{\mr{rot}}^{\mu\nu} -\sum_{l=2}^{\infty}\sum_{n=0}^{\infty}l [P_{\mce,n}^{[\mu}{}_{\mu_{L-1}}Q^{\nu]\mu_{L-1}}_{\mce,n}+(\mce\leftrightarrow \mcb)]$, to get
\begin{alignat}{3}
\label{seqn}
 \frac{D S^{\mu\nu}}{D\tau} &=  2 p^{[\mu}u^{\nu]}+\frac{4}{3}R^{[\mu}{}_{\lambda\rho\sigma}J_{\lambda}{}^{\nu]\tau\rho\sigma} \\& +\frac{2}{3}\nabla^{\lambda}R^{[\mu}{}_{\tau\rho\sigma}J_{\lambda}{}^{\nu]\tau\rho\sigma}+\frac{1}{6}\nabla^{[\mu}R_{\lambda\tau\rho\sigma}J^{\nu]\lambda\tau\rho\sigma}.\nnm
\end{alignat}
This is the appropriate definition of total spin angular momentum of the body as reinforced by the fact that this also shows up in the stress energy tensor (see Eq.~\ref{stress}) in the expected manner. To solve the equations of motion, we also need to impose a ``spin supplementary condition'' (SSC) to ensure that it is a spatial tensor with the right number of degrees of freedom. Here, we impose the ``covariant'' or Tulczyjew-Dixon SSC at the level of the equations of motion upon the total physical spin angular momentum as $S^{\mu\nu}p_{\mu}=0$. 

The equation of motion for momentum can be obtained by variation of the action with respect to the worldline $z^{\mu}(\tau)$, which can be done following the covariant approach as shown in Ref.~\cite{Marsat:2014xea} to get 
\begin{alignat}{3}
\label{peqn}
\frac{D p_{\mu}}{D\tau} &= -\frac{1}{2}R_{\mu\nu\rho\sigma}u^{\nu}S^{\rho\sigma}-\frac{1}{6}J^{\lambda\nu\rho\sigma}\nabla_{\mu}R_{\lambda\nu\rho\sigma}\\&-\frac{1}{12}J^{\tau\lambda\nu\rho\sigma}\nabla_{\mu}\nabla_{\tau}R_{\lambda\nu\rho\sigma}.\nnm
\end{alignat}
Note that only the sum of multipole moments $\sum_n Q_{\mce,n}^{\mu_l}$, appear in the expressions for the time derivatives of momentum and spin angular momentum. 

Finally, the stress energy tensor of the particle can be derived by varying the effective action with respect to the metric $g_{\mu\nu}$ as $T^{\mu\nu} = \frac{1}{\sqrt{-g}} \frac{\delta S}{\delta g_{\mu\nu}},$ where all dependency of the action on $g_{\mu\nu}$ needs to be taken into account during the variation. This variation was performed in Ref.~\cite{Marsat:2014xea} to obtain
\begin{widetext}
\begin{alignat}{3}
\label{stress}
T^{\mu \nu} & = T^{\mu \nu}_{\mr{pole-dipole}} + T^{\mu \nu}_{\mr{quadrupole}}+T^{\mu\nu}_{\mr{octupole}}, \\ T^{\mu \nu}_{pole-dipole} & = \int d\tau p^{(\mu} u^{\nu)} \frac{\delta^{(4)}(x-z)}{\sqrt{-g}} - \nabla_{\rho} \int d\tau S^{\rho (\mu} u^{\nu)}\frac{\delta^{(4)}(x-z)}{\sqrt{-g}}, \\
T^{\mu \nu}_{\mr{quadrupole}} & = \int d\tau \frac{1}{3} R^{(\mu}_{~\lambda \rho \sigma}J^{\nu) \lambda \rho \sigma}\frac{\delta^{(4)}(x-z)}{\sqrt{-g}} - \nabla_{\rho} \nabla_{\sigma}\int d\tau \frac{2}{3} J^{\rho (\mu \nu) \sigma} \frac{\delta^{(4)}(x-z)}{\sqrt{-g}},\\
T^{\mu \nu}_{\mr{octupole}} &= \int d\tau\Bigg[\frac{1}{6} \nabla^{\lambda} R^{(\mu}_{~\xi \rho \sigma}J_{\lambda}^{~~\nu) \xi \rho \sigma} + \frac{1}{12} \nabla^{(\mu} R_{\xi \tau \rho \sigma}J^{\nu) \xi \tau \rho \sigma}\Bigg]\frac{\delta^{(4)}(x-z)}{\sqrt{-g}} \\ & + \nabla_{\rho}\int d\tau\Bigg[-\frac{1}{6}R^{(\mu}_{~\xi \lambda \sigma}J^{|\rho|\nu)\xi\lambda\sigma} - \frac{1}{3}R^{(\mu}_{~\xi \lambda \sigma}J^{\nu)\rho \xi \lambda \sigma} + \frac{1}{3}R^{\rho}_{~\xi \lambda \sigma}J^{(\mu \nu)\xi \lambda \sigma}\Bigg]\frac{\delta^{(4)}(x-z)}{\sqrt{-g}} \\ &+ \nabla_{\lambda}\nabla_{\rho}\nabla_{\sigma} \int d\tau \frac{1}{3} J^{\sigma \rho (\mu \nu) \lambda}\frac{\delta^{(4)}(x-z)}{\sqrt{-g}},
\end{alignat}
\end{widetext}
which holds true even in the presence of inducible multipole moments when the contribution of the multipole moments is included in the definitions of the momentum and spin angular momentum. Since we are only interested in linear tides, we can drop all nonlinear (in curvature or metric perturbation) contributions in the quadrupolar and octupolar stress energy tensor to get a simpler truncated stress energy tensor as
\begin{widetext}
\begin{alignat}{3}
T^{\mu\nu} \nnm &= \int d\tau p^{(\mu} u^{\nu)} \delta^{(4)}(x-z) - \nabla_{\rho} \int d\tau S^{\rho (\mu} u^{\nu)}\delta^{(4)}(x-z) - \nabla_{\rho} \nabla_{\sigma}\int d\tau \frac{2}{3} J^{\rho (\mu \nu) \sigma} \delta^{(4)}(x-z)\\& + \nabla_{\lambda}\nabla_{\rho}\nabla_{\sigma} \int d\tau \frac{1}{3} J^{\sigma \rho (\mu \nu) \lambda}\delta^{(4)}(x-z).
\label{trstress}
\end{alignat}
\end{widetext}
This truncated stress energy tensor will be important later to solve the problem of GWs scattering off the effective particle.

 To proceed further, we need to relate the tidal fields to the multipole moments. We will now motivate and write down general ans{\"a}tze for the multipole moments.
\subsection{Ans{\"a}tze for multipole moments}
\label{nyayam}
 
 Going forward, we define the sum of all multipole moments for a given $l$ and parity (electric or magnetic) as individual moments for convenience as $ Q_\mce^{\mu_L}= \sum_n Q_{\mce,n}^{\mu_L}$. Now, as mentioned earlier, when the particle obeys spherical symmetry, only the tidal fields to which the multipole moment explicitly couples to in the action can affect it and we have Eq.~(\ref{scwmom}). However, in the presence of spin, it is possible to have more fields in the formula for multipole moments. Since we intend to model the spinning BH, which breaks spherical symmetry but still respects parity, we can only have tidal fields with the same transformation under parity as the moment in the right hand side. The transformation under reflection is $(-1)^l$ for $\mce$ fields and $(-1)^{l+1}$ for $\mcb$ fields. Furthermore, we restrict our attention to quadrupolar and octupolar tidal fields as the higher multipolar order tidal fields are not relevant to the order of interest in this work. Thus, we can write a general ansatz as
\begin{widetext}
\begin{alignat}{3}
Q^{\mu\nu}_{\mce} &= M\sum_m^{\mr{\infty}}(GM)^{4+m} \lambda_{\mce,m}^{\mu\nu}{}_{\rho\sigma}\frac{D^m}{D\tau^m}\mce^{\rho\sigma} + M\sum _m^{\infty}(GM)^{5+m}\zeta^{\mu\nu}_{\mcb,m}{}_{\rho\sigma\gamma}\frac{D^m}{D\tau^m}\mcb^{\rho\sigma\gamma}, \quad (\mce\leftrightarrow\mcb),\nonumber\\
Q_{\mcb}^{\mu\nu\rho} &= M\sum_m^{\infty}(GM)^{5+m}\eta^{\mu\nu\rho}_{\mce,m}{}_{\sigma\gamma}\frac{D^m}{D\tau^m}\mce^{\sigma\gamma} + M\sum_{m}^{\infty}(GM)^{6+m}\Lambda^{\mu\nu\rho}_{\mcb,m}{}_{\alpha\sigma\delta}\frac{D^m}{D\tau^m}\mcb^{\alpha\sigma\delta},\quad (\mce\leftrightarrow \mcb),
\label{rawans}
\end{alignat}
\end{widetext}
where we have rendered the response tensors dimensionless by removing factors of $GM$. The response tensors can only have a nontrivial structure due to the spin of the particle. Additionally, they must be orthogonal to $u^{\mu}$ and be traceless in the upper and lower set of indices separately. One can thus generally decompose the response tensors as linear combination of building-block tensors made up of the spin tensor $S^{\mu\nu}$, Pauli-Lubanski spin vector $s^{\mu}\approx-(1/2)\epsilon^{\mu}{}_{\nu\rho\sigma}u^{\nu}S^{\rho\sigma}$,\footnote{The Pauli Lubanski spin vector is actually defined as $s^{\mu} =- [1/(2m)]\epsilon^{\mu}{}_{\nu\rho\sigma}p^{\nu}S^{\rho\sigma}$. However, as $p^{\mu}=m u^{\mu} + \mathcal{O}(R)$, we can neglect the curvature-dependent corrections when substituting in the linear tidal-response.} and the orthogonal (to four-velocity $u^{\mu}$) projection  operator $\mathcal{P}^{\mu}_{\nu}=\delta^{\mu}_{\nu}+u^{\mu}u_{\nu}$. While infinitely many such combinations may be written, only  a handful of them are linearly independent and we can generally decompose the response tensors $\lambda,~\zeta,~\eta$, as (see, e.g., Ref.~\cite{Goldberger:2020fot} where the correlation function for the quadrupole moment was fixed)
\begin{widetext} 
\begin{alignat}{3}
\lambda_{\mce,n}^{\mu\nu}{}_{\rho\sigma} &= f_{\mce,n}^0\mathcal{P}^{\langle\mu}_{\langle\rho}\mathcal{P}^{\nu\rangle}_{\sigma\rangle}+f_{\mce,n}^{1}\hat{S}^{\langle\mu}{}_{\langle\rho}\mathcal{P}^{\nu\rangle}_{\sigma\rangle} + f_{\mce,n}^2 \hat{s}^{\langle\mu}\hat{s}_{\langle\rho}\delta^{\nu\rangle}_{\sigma\rangle} + f_{\mce,n}^3 \hat{s}^{\langle\mu}\hat{s}_{\langle\rho}\hat{S}^{\nu\rangle}{}_{\sigma\rangle}+f_{\mce,n}^4 \hat{s}^{\langle\mu}\hat{s}_{\langle\rho}\hat{s}^{\nu\rangle}\hat{s}_{\sigma\rangle},~(\mce\leftrightarrow \mcb) \nonumber \\
\zeta_{\mcb,n}^{\mu\nu}{}_{\rho\sigma\gamma}&= \lambda_{\mce,n}^{\mu\nu}{}_{\langle\rho\sigma}\hat{s}_{\gamma\rangle}(f_{\mce,n}\rightarrow g_{\mcb,n}), ~\eta_{\mce,n}^{\mu\nu\rho}{}_{\rho\sigma}=\hat{s}^{\langle\rho}\lambda_{\mce,n}^{\mu\nu\rangle}{}_{\rho\sigma}(f_{\mce,n}\rightarrow h_{\mce,n}),\quad (\mce\leftrightarrow \mcb) \label{formfixed}
\end{alignat}
\end{widetext}
where the sets of coefficients $f$, $g$, $h$ can now only depend on the spin parameter $\chi=J/(GM^2)$, where $J=\sqrt{(1/2)S^{\mu\nu}S_{\mu\nu}}$ is the magnitude of spin angular momentum. Also, we have defined $\hat{S}^{\mu\nu}$ and $\hat{s}^{\mu}$ as the normalized versions of the spin-tensor ($\hS^{\mu\nu}\hS_{\mu\nu} = 2$)  and spin-vector ($\hs^{\mu}\hs_{\mu}=1$) so they are dimensionless and independent of any parameter. Adding any other tensor made up of the same ingredients will be linearly dependent on the remaining pieces, which follows simply from the relations $\hS^{\mu\nu}\hs_{\nu}=0$, and $\hS^{\mu}{}_{\rho}\hS^{\nu}{}_{\sigma} = -\mathcal{P}^{\mu}{}_{\sigma}\mathcal{P}^{\nu}{}_{\rho}+\mathcal{P}^{\mu\nu}\mathcal{P}_{\rho\sigma}-\mathcal{P}_{\rho\sigma}\hs^{\mu}\hs^{\nu}+\mathcal{P}^{\mu}{}_{\sigma}\hs^{\nu}\hs_{\rho}+\mathcal{P}^{\nu}{}_{\rho}\hs^{\mu}\hs_{\sigma}-\mathcal{P}^{\mu\nu}\hs_{\rho}\hs_{\sigma}$, and thus no more free coefficients can be introduced. We have transferred all freedom in choosing the response to the associated sets of $\chi$-dependent parameters $f$, $g$, and $h$. We can similarly decompose the tensor $\Lambda^{\mu\nu\rho}_{\alpha\sigma\delta}$ but it turns out to be irrelevant to the order of interest in this work, so we just drop its contribution from now on. We can drop most of the time-derivatives of the tidal fields in the ans{\"a}tze for the same reason and work with simplified truncated ans{\"a}tze by including terms only up to $(GM)^5$,
\begin{alignat}{3}
 \label{trunc_moment}
\nnm Q_{\mce}^{\mu\nu} &= (GM)^4 \lambda_{\mce,0\rho\sigma}^{\mu\nu}\mce^{\rho\sigma} + (GM)^5 \lambda_{\mce,1\rho\sigma}^{\mu\nu}\frac{D}{D\tau} \mce^{\rho\sigma}\nnm  \\& + \nnm (GM)^5 \nu^{\mu\nu}_{\mcb,0}{}_{\langle\rho\sigma}\hat{s}_{\gamma\rangle}\mcb^{\rho\sigma\gamma},~(\mce\leftrightarrow \mcb),  \\
Q_{\mcb}^{\mu\nu\rho}&=(GM)^5 \hat{s}^{\langle\rho}\xi_{\mce,0}^{\mu\nu\rangle}{}_{\alpha\beta}\mce^{\alpha\beta},~(\mce\leftrightarrow \mcb),
\end{alignat}
where we have defined $\nu^{\mu\nu}_{\mcb,0}{}_{\rho\sigma} = \lambda^{\mu\nu}_{\mcb(\mce),0}{}_{\rho\sigma}(f_{\mce(\mcb),0}\rightarrow g_{\mcb(\mce),0})$, and $\xi^{\mu\nu}_{\mce,0}{}_{\rho\sigma} =\lambda^{\mu\nu}_{\mce,0}{}_{\rho\sigma}(f_{\mce(\mcb),0}\rightarrow h_{\mce,0})$. The 40 coefficients, $f_{\mce(\mcb),0}^i$, $f_{\mce(\mcb),1}^i$, $g_{\mcb(\mce),0}^i$, $h_{\mce(\mcb),0}$ characterize the tidal response of the particle. Some contribute to dissipative effects whereas others contribute only to conservative effects. Unlike for the spinless case, it is not obvious here which ones contribute to dissipation and which ones do not, due to coupling between different multipolar orders. At leading order, focussing only on the quadrupolar tidal field inducing the quadrupolar multipole moment, it can be shown (see Ref.~\cite{Goldberger:2020fot}) that the part of the tensor $\lambda^{\mu\nu\rho\sigma}_{\mce,0}$ that is antisymmetric under $\mu\nu \leftrightarrow \rho\sigma$ contributes to dissipation. This criteria is flipped for $\lambda^{\mu\nu}_{\mce,1\rho\sigma}$ as it is next to an odd power of a time-derivative. This includes the coefficients $f^{1}_{\mce(\mcb),0}$, $f^{3}_{\mce(\mcb),0}$, $f^{0}_{\mce(\mcb),1}$, $f^{2}_{\mce(\mcb),1}$, $f^{4}_{\mce(\mcb),1}$. It is less trivial for the coefficients entering tensors that mix the different multipolar orders. We will see in Sec.~\ref{doa} which coefficients are conservative and which are dissipative among these. We accomplish this by scattering GWs off the effective particle and computing the degree of absorption for the $l=2$ spheroidal modes and comparing with the analogous result  obtained by solving the Teukolsky equation. This comparison, along with demanding that the different spheroidal modes of the Newman-Penrose scalar $\psi_4$ scatter independently in the effective theory will help identify which coefficients lead to dissipation and which only to conservative effects. Additionally, we will also be able to fix the coefficients that contribute to dissipative effects through comparison with results from BHPT.
\section{Fixing the unknown parameters in ans{\"a}tze for the multipole moments}\label{sec:scatter}
To fix the dissipative part of the response, we now scatter GWs off the effective particle with mass $M$ and compute a suitable quantity indicating the degree of absorption. We can then fix the free parameters by comparison with the analogous quantity for actual spinning BHs, obtained by solving the Teukolsky equation. We will only consider the effect of the linearly induced tidal moments in the scattering of the wave in what follows, since the other contributions to the scattered wave are expected to be purely conservative or irrelevant (up to the order of interest). However, in this way we also neglect the leading-order effect of nonlinearities due to the wave scattering off the stationary gravitational background, leading to certain subtleties  when matching with the result from the Teukolsky equation which we will have to address.
\subsection{The GW environment}
Let the unperturbed particle be at the origin at rest, $u^{\mu}=(1,0,0,0)$ and $z^{\mu} = (\tau=t,0,0,0)$, where $\tau$ is the proper time of the particle which is identical to the background time $t$ when undisturbed. We add to the particle's background a general GW perturbation $h^{\mu\nu} = \sqrt{-g}g^{\mu\nu}-\eta^{\mu\nu}$. We choose the perturbation to be in the harmonic and transverse-traceless (TT) gauge, satisfying $\partial_{\mu}h^{\mu\nu}=0$, $h^{\mu\nu}u_{\nu}=0$, $h^{\mu\nu}\eta_{\mu\nu}=0$. At zeroth order (in $G$), when the interactions of the metric perturbation with the stationary gravitational field of the particle are neglected, it just satisfies the flat space-time wave equation $\eta^{\mu\nu}\partial_{\mu}\partial_{\nu}h^{\alpha\beta}=0$ (except at origin where the particle is present). Then, we can write the general tensor wave solution in the rest frame of the particle (defined by $u^{\mu}$) at leading order as
\begin{widetext}
\begin{alignat}{3}
\label{wav0sol}
h_{ij} = \sum_{l=2}^{\infty}\frac{1}{\omega^{l-2}}\bigg(C^{K_l}_{\mce,\mr{in}}\Pi_{ij}^{k_{l-1}k_{l}}\hat{\partial}_{K_{l-2}}+\frac{1}{\omega}C^{K_l}_{\mcb,\mr{in}}\Pi_{ij}^{k_l m}\epsilon_{k_{l-1}mn}\hat{\partial}_{K_{l-2}n}\bigg)\psi_{\mr{in}} + (\mr{in} \rightarrow \mr{out}),
\end{alignat}
\end{widetext}
 where $\omega$ is the frequency of the wave in the rest frame of the particle and $\Pi_{kl}^{ij}=(1/2)(P^{i}{}_{k}P^{j}{}_{l}+P^{j}{}_{k}P^{i}{}_{l}-P^{ij}{P}_{kl})$, with $P^{ij} = \delta^{ij} + \partial^{i}\partial^{j}/\omega^2$, is a differential projection operator which ensures that the harmonic and TT gauge conditions are  satisfied. $\psi_{\mr{in}}=\exp[-i \omega(t+r)]/(\omega r)$ and $\psi_{\mr{out}}=\exp[-i \omega(t-r)]/(\omega r)$ are the incoming and outgoing wave solutions for the $l=0$ mode of a scalar wave respectively. $C_{\mce/\mcb,\mr{in}/\mr{out}}^{K_{l}}$ are dimensionless symmetric trace-free (STF) tensors characterizing the amplitudes of each $l$ mode. $\mce$, $\mcb$ label the coefficients tuning the electric and magnetic polarizations respectively.\footnote{In general, there are two distinct solutions for each $\omega$, $l$, and $m$ for a massless tensor wave (except for scalars). We split them according to parity in this work, with $\mce$ modes transforming as $(-1)^l$ and $\mcb$ modes transforming as $(-1)^{l+1}$ under reflection. The labels are also related to the manner in which different coefficients contribute to the tidal fields (see Eq.~(\ref{fator0})).} The different $l$ modes and polarizations ($\mce$, $\mcb$) do not mix under rotations. However, that does not mean we can tune them separately in the presence of the particle, as the particle's inducible multipole moments combined with its spin can couple different modes with each other. The general solution in Eq.~(\ref{wav0sol}) can be understood by noting that a basis of solutions to the homogeneous wave equation in flat space-time (albeit allowing for irregular behaviour at the origin) can be obtained by acting arbitrary number of spatial derivatives upon the spherically symmetric solutions, i.e.,  $\psi_{\mr{in}}$, $\psi_{\mr{out}}$ and any linear combination of them. We can then generally write the solution for $h_{ij}$ as a linear combination of the basis solutions with undetermined tensors contracting them to get the appropriate tensor structure. One can then decompose them into rotationally independent pieces by splitting the undetermined tensors into STF tensors and then separating them according to parity ($\mce$ and $\mcb$). Finally, one uses the projection operator $\Pi^{ij}_{kl}$ to ensure $h^{ij}$ is traceless and satisfies the harmonic gauge condition.

  Now, we want to split the total general solution in Eq.~(\ref{wav0sol}) into two parts, one that can be regarded as the ``input'' part of the wave which corresponds to the part that induces the tidal moments and an ``output'' part which is sourced from the induced tidal moments. Naively, one might think that the input part can be obtained by simply setting the outgoing mode coefficients $C_{\mce/\mcb,\mr{out}}^{K_l}=0$, but that is incorrect as there will be an outgoing wave in general even in the absence of a particle due to the the fact that an incoming wave packet in the distant past becomes an outgoing wave packet in the distant future (after crossing the origin). Also, the incoming part of the wave by itself is irregular at the origin and thus cannot be sustained without the presence of a particle. In fact, the correct splitting is given by the regular (input) and irregular (output) parts of the wave respectively, as follows
\begin{widetext}
\begin{alignat}{3}
\label{regirr0}
h_{ij} &= \sum_{l=2}^{\infty}\frac{1}{\omega^{l-2}}\bigg(C^{K_l}_{\mce,\mr{reg}}\Pi_{ij}^{k_{l-1}k_{l}}\hat{\partial}_{K_{l-2}}+\frac{1}{\omega}C^{K_l}_{\mcb,\mr{reg}}\Pi_{ij}^{k_l m}\epsilon_{k_{l-1}mn}\hat{\partial}_{K_{l-2}n}\bigg)\psi_{\mr{reg}} + (\mr{reg} \rightarrow \mr{irr}),
\\
\psi_{\mr{reg}} &= \frac{1}{2 i}(\psi_{\mr{out}}-\psi_{\mr{in}}) = \frac{\exp(-i\omega t)\sin(\omega r)}{\omega r},\quad \psi_{\mr{irr}} = (\psi_{\mr{out}}+\psi_{\mr{in}}) = \frac{\exp(-i\omega t)\cos(\omega r)}{\omega r}, \\
C_{\mr{reg}} &= (C_{\mr{out}}-C_{\mr{in}}) i,\quad C_{\mr{irr}}=C_{\mr{out}}+C_{\mr{in}}, \label{rirr2inout}
\end{alignat}
\end{widetext}
where $\psi_{\mr{reg}}$ is regular at the origin and obeys the wave equation everywhere whereas $\psi_{\mr{irr}}$ is irregular at the origin and requires support from a source, such as the stress energy tensor of the effective particle. More specifically, $\eta^{\mu\nu}\partial_{\mu}\partial_{\nu} \psi_{\mr{reg}} = 0$, and $\eta^{\mu\nu}\partial_{\mu}\partial_{\nu} \psi_{\mr{irr}} = (4\pi/\omega)\delta^{(3)}(\vec{r})\exp(-i\omega t)$. Thus, we will use the regular part of $h_{ij}$, which has finite values at the origin to compute the tidal fields that will induce the multipole moments and the irregular part of the fields will be related to the stress energy tensor of the particle's induced tidal moments through the relation 
\begin{alignat}{3}
\label{dwaveqn}
\nnm \eta^{\mu\nu}\partial_{\mu}\partial_{\nu} h_{ij}^{\mr{irr}} &= \sum_{l=2}^{\infty}\frac{\exp(-i\omega t)}{\omega^{l-1}}\bigg(C^{K_l}_{\mce,\mr{irr}}\Pi_{ij}^{k_{l-1}k_{l}}\hat{\partial}_{K_{l-2}}\\&+\frac{1}{\omega}C^{K_l}_{\mcb,\mr{irr}}\Pi_{ij}^{k_l m}\epsilon_{k_{l-1}mn}\hat{\partial}_{K_{l-2}n}\bigg)4\pi \delta^{(3)}(\vec{r}) \nnm \\& = 16 \pi G|g|\Pi^{kl}_{ij}T_{kl}(\vec{r}), 
\end{alignat}
where we have projected out the stress energy tensor appropriately since we are focusing on the radiative part of the field and using $h_{ij}^{\mr{irr}}$ to label the irregular part of the metric perturbation (i.e., the part generated from acting spatial derivatives on $\psi_{\mr{irr}}$) . We will have to solve this relation with the stress energy tensor corresponding to that due to the induced quadrupolar and octupolar multipole moments given to leading order in the curvature tensor in Eq.~(\ref{trstress}).
\subsection{Induced multipole moments}
As mentioned earlier, the regular part of the wave, which can exist without support and satisfies the homogeneous wave equation,  should be seen as the input part of the wave. We will use this part of the metric perturbation to compute the tidal fields which will induce the multipole moments. With the definitions given earlier in Eqs.~(\ref{tflds}), and the formula for the regular part of the wave given in Eq.~(\ref{regirr0}), we find that the value of the tidal fields at the origin is given by
\begin{subequations}
\label{fator0}
\begin{alignat}{3}
\mce^{ij}_{\mr{origin}} &= -\frac{1}{5}\omega^2 C^{ij}_{\mce,\mr{reg}},\\
\mcb^{ij}_{\mr{origin}} &= \frac{i}{5}\omega^2C^{ij}_{\mcb,\mr{reg}},  \\
\mce^{ijk}_{\mr{origin}}&= \frac{1}{21}\omega^3 C^{ijk}_{\mce,\mr{reg}}, 
\\ \mcb^{ijk}_{\mr{origin}} &= \frac{-i}{21}\omega^3 C^{ijk}_{\mcb,\mr{reg}}.
\end{alignat}
\end{subequations}

The multipole moments can now be computed by substituting these fields in Eqs.~(\ref{trunc_moment}). However, we see that the nontrivial (due to spin) response tensors mix the various components together. While this is inconvenient, a simple way to rewrite these expressions in a basis that does not mix components is to orient the coordinate system so that the Pauli-Lubanski spin vector is along the z-axis and expand in spin-weighted spherical harmonics. For later convenience, we choose spherical harmonics with spin weight `-2'. The simplest way to transform to this basis from the Cartesian basis is to define
\begin{alignat}{3}
\label{moctet}
&\nonumber\vec{m}= \frac{1}{\sqrt{2}}(\hat{\theta}+i \hat{\phi}), \\& \nnm
\hat{\theta} = \cos(\theta)\cos(\phi)\hat{x}+\cos(\theta)\sin(\phi)\hat{y}-\sin(\theta)\hat{z}, \\& \nnm
\hat{\phi}= -\sin(\phi)\hat{x}+\cos(\phi)\hat{y}, \\
&\hat{r}=\sin(\theta)\cos(\phi)\hat{x}+\sin(\theta)\sin(\phi)\hat{y}+\cos(\theta)\hat{z},
\end{alignat}
and then expand $Q^{ij}\bar{m}_{i}\bar{m}_{j}$ and $O^{ijk}\bar{m}_{i}\bar{m}_{j}\hat{r}_{k}$ in spin weight -2, spherical harmonics by projection as
\begin{alignat}{3}
\label{spbmom}
\nnm &Q^{l=2,m}_{\mce/\mcb} = \int d\Omega~ Q^{ij}_{\mce/\mcb}\bar{m}_{i}\bar{m}_{j} \bar{Y}^{l=2,m}_{-2}(\theta,\phi), \\& O^{l=3,m}_{\mce/\mcb} = \int d\Omega~ Q^{ijk}_{\mce/\mcb}\bar{m}_{i}\bar{m}_{j}\hat{r}_{k} \bar{Y}^{l=3,m}_{-2}(\theta,\phi),
\end{alignat}
where we are using $Y_{s=-2}^{lm}(\theta,\phi)$ to denote spherical harmonics with spin weight -2. They are normalized to 1 [i.e., $\int d\Omega ~Y_{-2}^{lm}(\theta,\phi)\bar{Y}_{-2}^{lm}(\theta,\phi)=1$]. We similarly define the incoming and outgoing mode coefficients in this basis as
\begin{alignat}{3}
\nnm
\label{spbmc}
C^{l=2,m}_{\mce/\mcb} &= \int d\Omega ~C^{ij}_{\mce/\mcb}\bar{m}_{i}\bar{m}_{j} \bar{Y}_{-2}^{2m}(\theta,\phi) \\ C^{l=3,m}_{\mce/\mcb} &=\int d\Omega~ C^{ijk}_{\mce/\mcb}\bar{m}_{i}\bar{m}_{j}\hat{r}_{k}\bar{Y}_{-2}^{3m}(\theta,\phi),
\end{alignat} 
where we have suppressed the `$\mr{in}$' (`$\mr{out}$') subscript for brevity. The definitions are identical for both incoming and outgoing mode coefficients.
Note that we are using the symbol $O$ for octupole tensor in spherical harmonic (with spin weight -2) basis. Then, we can write down the expressions for the multipole moments obtained by substituting Eq.~(\ref{fator0}) into the ans{\"a}tze in Eq.~(\ref{trunc_moment}) simply as 
\begin{widetext}
\begin{alignat}{3}
\label{lmmom}
Q^{2m}_{\mce} &= -e^{-i\omega t}M\omega^2\frac{(GM)^4}{30}\{6\mathcal{F}^{0,\mr{reg}}_{\mce}+ 3i\mathcal{F}_{\mce}^{1,\mr{reg}}~m+(m^2-4)[-\mathcal{F}^{2,\mr{reg}}_{\mce}-i\mathcal{F}^{3,\mr{reg}}_{\mce}~m+\mathcal{F}^{4,\mr{reg}}_{\mce}~(m^2-1)]\}, \nonumber\\
O^{3m}_{\mcb} &= -e^{-i\omega t}M\omega^2\frac{\sqrt{9-m^2}(GM)^5}{90\sqrt{7}}\{6h^{0}_{\mce}+ 3ih^1_{\mce}~m+(m^2-4)[-h^2_{\mce}-ih^3_{\mce}~m+h^4_{\mce}~(m^2-1)]\}C^{2m}_{\mce,\mr{reg}},\nonumber \\Q^{2m}_{\mcb} &= ie^{-i\omega t}M\omega^2\frac{(GM)^4}{30}\{6\mathcal{F}^{0,\mr{reg}}_{\mcb}+ 3i\mathcal{F}_{\mcb}^{1,\mr{reg}}~m+(m^2-4)[-\mathcal{F}^{2,\mr{reg}}_{\mcb}-i\mathcal{F}^{3,\mr{reg}}_{\mcb}~m+\mathcal{F}^{4,\mr{reg}}_{\mcb}~(m^2-1)]\},\nnm \\O^{3m}_{\mce} &= ie^{-i\omega t}M\omega^2\frac{\sqrt{9-m^2}(GM)^5}{90\sqrt{7}}\{6h^{0}_{\mcb}+ 3ih^1_{\mcb}~m+(m^2-4)[-h^2_{\mcb}-ih^3_{\mcb}~m+h^4_{\mcb}~(m^2-1)]\}C^{2m}_{\mcb,\mr{reg}},
\end{alignat}
\end{widetext}
where $\mathcal{F}^{i,\mr{reg}}_{\mce(\mcb)} =  (f^{i}_{\mce(\mcb),0}-iGM\omega f^{i}_{\mce(\mcb),1})C^{2m}_{\mce(\mcb),\mr{reg}}+[(i\sqrt{9-m^2})/(3\sqrt{7})]GM\omega g^{i}_{\mcb(\mce),0}C^{3m}_{\mcb(\mce),\mr{reg}}$, and we see that there is no longer any mixing of different $m$ modes. This is simply because we oriented the coordinate system so that the z-axis is along the spin, and thus it has 0 azimuthal quantum number to add or remove from that of the STF tensors characterizing the field. However, there is still mixing between different $l$ modes due to the tidal response mixing the quadrupolar and octupolar sectors. This complicates the process of defining a degree of absorption or a scattering phase, and we will  tackle this problem later in Sec.~\ref{doa}, by switching to a basis where there is no mixing of modes.
\subsection{Solving for the outgoing wave}
Now, we can use the wave equation with the appropriate source in Eq.~(\ref{dwaveqn}) to relate the multipole moments to the irregular part of the wave. The relevant (projected) part of the stress energy tensor in the chosen coordinate system is given by
\begin{alignat}{3}
\nnm T^{kl}\Pi^{ij}_{kl} &= \Pi^{ij}_{kl}\Big(\frac{1}{2}\ddot{Q}_\mce^{kl} +\frac{1}{2}\epsilon^{\langle k}{}_{mn}\dot{Q}_\mcb^{l\rangle n}\partial^{m}-\frac{1}{2}\ddot{Q}_\mce^{kl}{}_m\partial^m \\& -\frac{1}{2}\epsilon^{\langle k}{}_{mn}\dot{Q}_\mcb^{l\rangle no}\partial^{m}\partial_o\Big)\delta^{(3)}(\vec{x}), \label{projstr}
\end{alignat}
which is obtained by substituting the expressions for $J^{\mu\nu\rho\sigma}$ and $J^{\lambda\mu\nu\rho\sigma}$ in terms of the multipole moments given in Eqs.~(\ref{4quad}), (\ref{4oct}) into the stress energy tensor Eq.~(\ref{trstress})  and discarding terms with components along $u^{\mu}$ (as they will be eliminated upon projection). Now, substituting Eq.~(\ref{projstr}) into the RHS of Eq.~(\ref{dwaveqn}) and comparing, we get the relations
\begin{alignat}{3}
e^{-i\omega t}C^{ij}_{\mce,irr}&=2 G \omega^3 Q_\mce^{ij},~e^{-i\omega t}C^{ij}_{\mcb,irr} = 2 i G\omega^3 Q^{ij}_\mcb,\nonumber \\ e^{-i\omega t}C^{ijk}_{\mce,irr}&=-2G \omega^4 Q_\mce^{ijk},~e^{-i\omega t}C^{ijk}_{\mcb,irr}=-2iG\omega^4 Q_\mcb^{ijk},\nnm \\ & \label{mom2irr}
\end{alignat}
which are simple proportionality relations and thus can be trivially transformed to the $l,m$ basis.

 Substiting Eq.~(\ref{mom2irr}) in Eq.~(\ref{lmmom}), we get the relations
\begin{widetext}
\begin{alignat}{3}
C^{2m}_{\mce/\mcb,\mr{irr}} &= -\frac{\epsilon^5}{15}\{6\mathcal{F}^{0,\mr{reg}}_{\mce/\mcb}+ 3i\mathcal{F}_{\mce/\mcb}^{1,\mr{reg}}~m+(m^2-4)[-\mathcal{F}^{2,\mr{reg}}_{\mce/\mcb}-i\mathcal{F}^{3,\mr{reg}}_{\mce/\mcb}~m+\mathcal{F}^{4,\mr{reg}}_{\mce/\mcb}~(m^2-1)]\}, \nonumber\\\nonumber 
C^{3m}_{\mcb,\mr{irr}} &= i\frac{\sqrt{9-m^2}\epsilon^6}{45\sqrt{7}}\{6h^{0}_{\mce}+ 3ih^1_{\mce}~m+(m^2-4)[-h^2_{\mce}-ih^3_{\mce}~m+h^4_{\mce}~(m^2-1)]\}C^{2m}_{\mce,\mr{reg}},\nonumber\\
C^{3m}_{\mce,\mr{irr}} &=-i\frac{\sqrt{9-m^2}\epsilon^6}{45\sqrt{7}}\{6h^{0}_{\mcb}+ 3ih^1_{\mcb}~m+(m^2-4)[-h^2_{\mcb}-ih^3_{\mcb}~m+h^4_{\mcb}~(m^2-1)]\}C^{2m}_{\mcb,\mr{reg}},
\end{alignat}
 \end{widetext} 
 where we have defined $\epsilon=GM\omega$.
 Now, we can use the relations in Eqs.~(\ref{rirr2inout}) to solve the the outgoing coefficients. Since we are only interested in linear tides, and the leading-order tidal effects already start at a very high order in $\epsilon$, we can approximate in the RHS as $C^{lm}_{\mce/\mcb,\mr{reg}}\approx -2iC^{lm}_{\mce/\mcb,\mr{in}} + \mathcal{O}(\epsilon^5)$. Then we get the expressions for $C^{lm}_{\mce/\mcb,\mr{out}}$ as 
\begin{widetext}
\begin{alignat}{3}
\nnm
C^{2m}_{\mce(\mcb),\mr{out}} &= -C^{2m}_{\mce(\mcb),\mr{in}} + \frac{2 i \epsilon^5}{15}\{6\mathcal{F}^{0,\mr{in}}_{\mce(\mcb)}+ 3i\mathcal{F}_{\mce(\mcb)}^{1,\mr{in}}~m+(m^2-4)[-\mathcal{F}^{2,\mr{in}}_{\mce(\mcb)}-i\mathcal{F}^{3,\mr{in}}_{\mce(\mcb)}~m+\mathcal{F}^{4,\mr{in}}_{\mce(\mcb)}~(m^2-1)]\}, \\
C^{3m}_{\mcb,\mr{out}} &= -C^{3m}_{\mcb,\mr{in}}+\frac{2\sqrt{9-m^2}\epsilon^6}{45\sqrt{7}}\{6h^{0}_{\mce}+ 3ih^1_{\mce}~m+(m^2-4)[-h^2_{\mce}-ih^3_{\mce}~m+h^4_{\mce}~(m^2-1)]\}C^{2m}_{\mcb,\mr{in}},\nonumber\\
C^{3m}_{\mce,\mr{out}} &= -C^{3m}_{\mce,\mr{in}}-\frac{2\sqrt{9-m^2}\epsilon^6}{45\sqrt{7}}\{6h^{0}_{\mcb}+ 3ih^1_{\mcb}~m+(m^2-4)[-h^2_{\mcb}-ih^3_{\mcb}~m+h^4_{\mcb}~(m^2-1)]\}C^{2m}_{\mcb,\mr{in}}.
\label{in2out}
\end{alignat}
\end{widetext}
\subsection{Degree of absorption}
\label{doa}
Now we need to compute a suitable quantity that measures the degree of absorption. If there were no spin, there would be no mixing of different $l$ modes, and we could simply define the degree of absorption (or emission) for each $l,m$ mode as $1-|C^{lm}_{\mce/\mcb,\mr{out}}/C^{lm}_{\mce/\mcb,\mr{in}}|$. We can still do the same if we instead use a different basis, formed by a linear combination of the mode coefficients in the spherical basis. To guess the appropriate combination (basis) in which the modes should scatter independently, we now turn to hints from BHPT. In BHPT, the Teukolsky equation \cite{Teukolsky:1972my} governs the behaviour of curvature perturbations in an exact Kerr background. Specifically, it governs the behaviour of the spin weight (-2) Teukolsky scalar ${}_{-2}\psi(t,r,\theta\phi)$ related to the standard Newman-Penrose curvature scalar $\psi_4$ by ${}_{-2}\psi = (r-ia \cos(\theta))^4 \psi_4$, where $a=GM\chi$.  A crucial property of the Teukolsky equation for ${}_{-2}\psi$ is that it is separable in spheroidal harmonic basis with spin weight -2 with fixed frequency eigen solutions with the form ${}_{-2}\psi \propto \exp(-i\omega t)S^{lm}_{-2}(\theta,\phi,a\omega){}_{-2}R_{lm\omega}(r)$ where we are using $S_{s=-2}^{lm}$ to denote normalized (to 1) spheroidal harmonics with spin weight -2 and ${}_{-2}R_{lm\omega}(r)$ is an eigen solution to the radial Teukolsky equation (see Sec.~\ref{match}). Asymptotically far away from the BH, i.e., as $r\rightarrow \infty$, the standard Newman-Penrose curvature scalar $\psi_4$ takes the form (see Eq.~(2.6) in Ref.~\cite{1978ApJS...36..451M})
\begin{widetext}
\begin{alignat}{3}
\label{intruth}
\psi_4 \rightarrow\; & \omega^2 \sum_{l,m,P=\pm 1} \Big\{K_{lm P}^{\mr{out}} \frac{\exp[-i\omega (t-r^{*})]}{\omega r} + \frac{1}{16}[\mr{Re}(C)+ 12 i \epsilon P]\frac{1}{\omega^4r^4}K_{lmP}^{\mr{in}} \frac{\exp[-i\omega(t+r^{*})]}{\omega r}\Big\}S^l_{-2,m}(\theta,\phi)  \nnm \\&+ \mathcal{O}\bigg(\frac{1}{r^6}\bigg),
\end{alignat}
\end{widetext}
where $r^{*}$ is the tortoise coordinate and $P=\pm 1$ is the index denoting the transformation under parity for each mode with $P=1$ ($P=-1$) for modes that are symmetric (anti-symmetric) under reflection. The expression for $\mr{Re}(C)$ can be found in Eq.~(31) of Ref.~\cite{Dolan:2008kf}. The outgoing and incoming mode coefficients for each spheroidal mode $l,m$ and parity $P=\pm 1$ are related simply as
\begin{alignat}{3}
\label{truphase}
K^{\mr{out}}_{lmP}  =  -(-1)^{l+m}\eta_{lm}^P\exp(2i\delta_P^{lm}) K^{\mr{in}}_{lmP}, 
\end{alignat}
where $\delta_P^{lm}\in \mathcal{R}$ is the conservative scattering phase for each mode and $\eta_{lm}$ is the degree of absorption/emission (i.e., the mode coefficients do not mix under scattering). Although these scattering phases and degree of absorption are defined for these abstract mode coefficients $K^{\mr{out}/\mr{in}}_{lmP}$, they are directly related to the scattering of a GW off a Kerr BH and often employed for that purpose (see for e.g., Refs.~\cite{Dolan:2007ut, Dolan:2008kf}). In particular, $\eta_{lm}^P$ characterizes the dissipation due to energy flux into the horizon with the transmission factor for each mode defined as  
\begin{alignat}{3}
T_{lm} = 1- |\eta_{lm}^{\pm}|^2.
\end{alignat}
Absorption or emission of energy at the horizon is only nonzero when $|\eta_{lm}^P|$ differs from 1, and thus this is a suitable quantity to be labelled as "degree of absorption" in the real theory.

In the effective worldline theory, we can continue to use the same quantity as the degree of absorption provided we relate the outgoing (incoming) mode coefficients $C^{lm}_{\mce/\mcb,\mr{out}(\mr{in})}$ of the wave-like metric perturbation to the outgoing (incoming) mode coefficients of $\psi_4$ i.e.,  $K^{\mr{out}(\mr{in})}_{lmP}$. In other words, relating $C^{lm}_{\mce/\mcb,\mr{out}(\mr{in})}$ to $K^{\mr{out}(\mr{in})}_{lmP}$ will reveal the basis in which the modes of the wave will scatter without mixing. We accomplish this by writing down the asymptotic behaviour of $\psi_4$ in the effective theory using its definition $\psi_4=-R_{\mu\nu\gamma\delta}\bar{m}^{\mu}\bar{m}^{\gamma}n^{\nu}n^{\delta}$, where $m^{\mu}$, $\bar{m}$, $n^{\mu}$, $l^{\mu}$ form a null tetrad field satisfying $l^2=m^2=n^2=0=l\cdot m=n\cdot m$, $l\cdot n= -1, m\cdot \bar{m}=1$, where $l^{\mu}$ and $n^{\mu}$ are real vectors whereas $m^{\mu}$ is a complex null vector with $\bar{m}^{\mu}$ as its complex conjugate. In flat space, or at leading order when the background curvature may be treated as a small perturbation such as far away from the particle, we can write them in term of spherical polar coordinates as $\hat{m}=(\hat{\theta}+i\hat{\phi})/\sqrt{2}$, $n^{\mu} = (t^{\mu}-r^{\mu})/\sqrt{2}$, $l^{\mu} = (t^{\mu}+\hat{r}^{\mu})/\sqrt{2}$. In the effective worldline picture, we can thus evaluate $\psi_4$ at large distances by using the flat space-time null tetrad, using the linearized curvature due to the wave-like metric perturbation in Eq.~(\ref{wav0sol}), yielding
\begin{alignat}{3}
\psi_4 &= \omega^2\frac{\exp[-i\omega (t-r)]}{\omega r}\\&\times\sum_{a=\mce,\mcb}\sum_{l=2}^{\infty}i^{l-2}C_{a,out}^{K_{l-2}ij}\hat{r}_{K_{l-2}}\hat{m}_{i}\hat{m}_{j} +\mathcal{O}\Big(\frac{1}{r^2}\Big) \nnm,
\end{alignat}
which we can rewrite as an expansion in spherical harmonics of spin weight -2 using the conventions in Eq.~(\ref{spbmc}) as
\begin{alignat}{3}
\psi_4 &=\omega^2\frac{\exp[-i\omega (t-r)]}{\omega r}\sum_{a=\mce,\mcb}[\sum_{m=-2}^{2}Y^{2m}_{-2}(\theta,\phi)C_{a,\mr{out}}^{lm} \nnm \\&+i \sum_{m=-3}^3 Y^{3m}_{-2}(\theta,\phi)C_{a,\mr{out}}^{lm}]+\mathcal{O}\Big(\frac{1}{r^2}\Big), 
\end{alignat}
where we have dropped modes above $l=3$ (hexadecapolar and above) as they are not relevant to the order of interest here.  Also note that that our effective theory does not couple $l\geq 4$ modes with any of the lower modes (up to the order of interest in this work) and thus we can safely set the associated STF tensors to zero.

We can then switch to spheroidal harmonics of spin weight -2 via the relations $Y^{2m}_{-2}=S^{2m}_{-2}-\epsilon\chi(2\sqrt{9-m^2})/(9\sqrt{7})S^{3m}_{-2}$, $Y^{3m}_{-2}=S^{3m}_{-2}+\epsilon\chi(2\sqrt{9-m^2})/(9\sqrt{7})S^{2}_{-2,m}$, valid at leading order in spheroidicity $=\epsilon\chi$, which is sufficient for our purposes. Also note that we have dropped the contribution to $Y^{3m}_{-2}$ from $S_{-2}^{l=4,m}$. This gives us

\begin{alignat}{3}
\label{inweft} \psi_4 &= \omega^2 \frac{\exp[-i\omega(t-r)]}{\omega r} \sum_{l,m,P}^{3}\mathcal{C}^{lm}_{P,\mr{out}}S_{-2}^{lm}(\theta,\phi)\nnm \\& + \mathcal{O}\bigg(\frac{1}{r^2}\bigg),
\end{alignat}
where we have defined the coefficients characterizing spheroidal modes as
\begin{widetext}
\begin{alignat}{3}
\label{sp2sph2}
\mathcal{C}^{2m}_{P=1,\mr{out}} &= \Big(C^{2m}_{\mce,\mr{out}}+i\frac{2\epsilon\chi\sqrt{9-m^2}}{9\sqrt{7}}C^{3m}_{\mcb,\mr{out}}\Big),
~\mathcal{C}^{2m}_{P=-1,\mr{out}} =\Big(C^{2m}_{\mcb,\mr{out}}+i\frac{2\epsilon\chi\sqrt{9-m^2}}{9\sqrt{7}}C^{3m}_{\mce,\mr{out}}\Big),\\
\mathcal{C}^{3m}_{P=1,\mr{out}} &=\Big(iC_{\mcb,\mr{out}}^{3m}-\frac{2\epsilon\chi\sqrt{9-m^2}}{9\sqrt{7}}C^{2m}_{\mce,\mr{out}}\Big)  ,
~\mathcal{C}^{3m}_{P=-1,\mr{out}} =\Big(iC_{\mce,\mr{out}}^{3m}-\frac{2\epsilon\chi\sqrt{9-m^2}}{9\sqrt{7}}C^{2m}_{\mcb,\mr{out}}\Big) ,
\label{sp2sph3}
\end{alignat}
\end{widetext}
and $P$ is used to split them according to parity, with $P=1$ for parity symmetric modes and $P=-1$ for parity antisymmetric modes. We can now compare with the known asymptotic form of $\psi_4$ from Eq.~(\ref{intruth}) to leading order in $1/r$. This gives us the simple identification between the coefficients of the outgoing modes of $\psi_4$ and that of the GW in effective theory as
\begin{alignat}{3}
\label{truph}
K^{\mr{out}}_{lmP} = \mathcal{C}^{lm}_{P,\mr{out}},
\end{alignat}
where we used the fact that the tortoise coordinate $r^*$ asymptotes to $r$ in the limit $r\rightarrow \infty$. Now, we know the appropriate combination of coefficients of outgoing modes in the effective theory to use. To get a similar relation between $K^{\mr{in}}_{lmP}$ and the coefficients of incoming modes in the effective theory, the simplest way is to just guess the form by considering the limit where there is no scattering. In the real theory, this is the limit where $\eta_{lm}^P=1$ and $\delta_P^{lm}=0$, and we have $K^{\mr{up}}_{lmP}=-(-1)^{l+m}K^{\mr{in}}_{lmP}$. In the effective theory, this is simply when the irregular part of the wave should vanish or $C_{\mce/\mcb,\mr{irr}}^{lm}=0\implies C_{\mce/\mcb,\mr{in}}^{lm} = -C_{\mce/\mcb,\mr{in}}^{lm}$, which in turn implies $\mathcal{C}_{P=\pm 1,\mr{out}}^{2m} = -\mathcal{C}_{P=\pm 1,\mr{in}}^{2m}$, where $\mathcal{C}_{P=\pm 1,\mr{in}}^{2m}$ is defined exactly as $\mathcal{C}_{P=\pm 1,\mr{out}}^{2,m}$ in Eqs.~(\ref{sp2sph2}),(\ref{sp2sph3}) except after transforming $\mr{out}\rightarrow \mr{in}$. Thus, we can simply identify $K^{\mr{in}}_{lmP} = (-1)^{l+m}\mathcal{C}_{P=\pm 1,\mr{in}}^{2,m}$. Then, the scattering phase relation in Eq.~(\ref{truphase}) can now be written in the effective theory simply as
\begin{alignat}{3}
\label{trit}
\mathcal{C}^{lm}_{P,\mr{out}} = -\eta^P_{lm}\exp(2i\delta_P^{lm})\mathcal{C}^{lm}_{P,\mr{in}},
\end{alignat}
and the degree of absorption can be defined in the effective worldline theory as 
\begin{alignat}{3}
\label{eft2realabs}
1-\eta^P_{lm}=1-\bigg|\frac{\mathcal{C}^{lm}_{P,\mr{out}}}{\mathcal{C}^{lm}_{P,\mr{in}}}\bigg|.
\end{alignat}
Note that in this way, we have defined a common quantity as the degree of absorption valid for both real and effective setups. Thus, this quantity will also serve for matching between the real and effective theories to fix the unknown tidal coefficients in the next subsection. 

Essentially, comparing the form of $\psi_4$ in the full and effective theories at large distances has revealed to us the combination of incoming and outgoing coefficients in the spherical basis that scatter without mixing. However, unsurprisingly, for general choices of tidal coefficients, the relation between the outgoing and incoming spheroidal coefficients is not going to nicely factorize as given in Eq.~(\ref{trit}). In fact, the tidal coefficients that mix the spherical $l=2$, and $l=3$ modes $g^{i}_{\mce/\mcb}$ and $h^i_{\mce/\mcb}$ have to be chosen on the effective theory side such that the combinations given in Eqs.~(\ref{sp2sph2}), (\ref{sp2sph3}) scatter without mixing. In this work, we impose this on the effective theory by plugging in the expressions for the spherical outgoing modes from Eq.~(\ref{in2out}) into the LHS of Eq.~(\ref{trit}) and demanding that it be proportional to the RHS, i.e., $\mathcal{C}^{lm}_{P,\mr{in}}$. We refer to this as imposing spheroidal separability on the effective theory and doing so yields the constraints 
 
\begin{alignat}{3}
\label{const}
\nnm g_{\mcb,0}^{i} &= h_{\mce,0}^i = \frac{2}{3}\chi f_{\mce,0}, \\ g_{\mce,0}^{i} &=h_{\mcb,0}^i = -\frac{2}{3}\chi f_{\mcb,0}^i, 
\end{alignat}
which greatly reduces our list of unknown variables entering the tidal response, and fixes the ratio of the coefficients connecting the octupole(quadrupole) fields to the quadrupole(octupole) moments, i.e., the coefficients $g^{i}_{\mce/\mcb,0}$($h^{i}_{\mce/\mcb,0}$), to the coefficients connecting the quadrupole fields to quadrupole moments, i.e., the coefficients $f^{i}_{\mce/\mcb,0}$.

Provided these relations are true, The different spheroidal modes will be scattered without mixing by the tidal moments in the effective worldline theory and we can compute the degree of absorption for the various modes using Eq.~(\ref{eft2realabs}) with constraints in Eq.~(\ref{const}) to be\footnote{We only focus on the degree of absorption and not on the conservative phase as the conservative phase also gains contributions from effects other than tidally induced multipole moments in the effective theory, for e.g., spin-induced multipole moments. A matching of the conservative phase can only be performed when all such effects are included in the solution of the scattering problem in the effective theory, which we do not.}

\begin{widetext}
\begin{alignat}{3}
1-\bigg|\frac{\mathcal{C}^{2m}_{P=1,\mr{out}}}{\mathcal{C}^{2m}_{P=1,\mr{in}}}\bigg| &=-\epsilon^5 m\bigg[\frac{2 f_{\mce,0}^1}{5}-(m^2-4)\frac{2}{15}f_{\mce,0}^3\bigg] +\epsilon^6\bigg\{\frac{4 f^0_{\mce,1} }{5}-(m^2-4)\bigg[\frac{2 f^2_{\mce,1}}{15}+\frac{2 f^4_{\mce,1}}{15}(m^2-1)\bigg]\bigg\}\nnm \\& \quad + \mathcal{O}(\epsilon^7) ,\nnm \\
1-\bigg|\frac{\mathcal{C}^{2m}_{P=-1,\mr{out}}}{\mathcal{C}^{2m}_{P=-1,\mr{in}}}\bigg| &=-\epsilon^5 m\bigg[\frac{2 f_{\mcb,0}^1}{5}-(m^2-4)\frac{2}{15}f_{\mcb,0}^3\bigg] +\epsilon^6\bigg\{\frac{4 f^0_{\mcb,1} }{5}-(m^2-4)\bigg[\frac{2 f^2_{\mcb,1}}{15}+\frac{2 f^4_{\mcb,1}}{15}(m^2-1)\bigg]\bigg\}\nnm \\& \quad + \mathcal{O}(\epsilon^7) ,\nnm \\
1-\bigg|\frac{\mathcal{C}^{3m}_{P=1,\mr{out}}}{\mathcal{C}^{3m}_{P=1,\mr{in}}}\bigg| &=1-\bigg|\frac{\mathcal{C}^{3m}_{P=-1,\mr{out}}}{\mathcal{C}^{3m}_{P=-1,\mr{in}}}\bigg| =  \mathcal{O}(\epsilon^7),
\label{weftabs}
\end{alignat}	
\end{widetext}
where we find that the degree of absorption for the spheroidal $l=3$ modes vanishes up to $\mathcal{O}(\epsilon^7)$ in the effective theory. Also note that the coefficients $f^0_{\mce/\mcb,0}$, $f^2_{\mce/\mcb,0}$, $f^4_{\mce/\mcb,0}$, $f^1_{\mce/\mcb,1}$, $f^3_{\mce/\mcb,3}$ do not appear anywhere in the degree of absorption. These coefficients thus only add to the conservative phase, and we will see later in Eqs.~(\ref{expdm}), (\ref{expdJ}) that they behave similarly with mass/spin evolution equations as well contributing only total time derivatives to $dm/dt$ and $dJ/dt$. We can now compare Eq.~(\ref{weftabs}) with the degree of absorption derived in the real theory by solving the Teukolsky equation for $\eta^P_{lm}$ and fix the coefficients that contribute to dissipation. However, because we solved the scattering problem in effective theory in flat space-time thus ignoring the nonlinear interactions between the wave and the stationary gravitatonal field of the particle, there are some subtleties in this matching process which we will have to tackle. We will very briefly outline the computation of $\eta^P_{lm}$ in the real theory by solving the Teukolsky equation and then match with the result from the effective theory while keeping the subtleties in mind.

\subsection{Matching with Teukolsky solution}
\label{match}
In the full theory, the degree of absorption defined from the $\mathcal{O}(1/r^6)$ expansion of $\psi_4$ in Eq.~(\ref{intruth}) as $1-|K^{out}_{lmP}/K^{in}_{lmP}|$ can be computed from analytical solutions \cite{Sasaki:2003xr} to the Teukolsky equation \cite{Teukolsky:1972my} as follows. Following the review \cite{Sasaki:2003xr}, the Teukolsky equation for ${}_{-2}\psi$ [$=(r-ia\cos(\theta))^4\psi_4$ in Boyer-Lindquist coordinates] is separable with fixed-frequency solutions given by ${}_{-2}\psi\propto e^{-i\omega t}\,S^{lm}_{s}(\theta,\phi,a\omega)\,{}_{-2}R_{lm\omega}(r)$, where $S^{lm}_{s}(\theta,\phi,a\omega)$ are spin-weighted spheroidal harmonics, and ${}_{s}R_{lm\omega}(r)$ is a solution to the (homogeneous) radial Teukolsky equation, with $s=-2$,
\begin{alignat}{3}
&\bigg[\Delta^{-s}\frac{d}{dr}\left(\Delta^{s+1}\frac{d}{dr}\right)+\frac{K^2-2 i s (r-GM)K}{\Delta}
\nonumber\\
&\quad+4 i s \omega r -{}_s \lambda_{lm}\bigg]{}_s R_{\ell m \omega}(r)
=0, \label{Eq:RadTeuk} 
\end{alignat}
where $\Delta=r^2+a^2-2GMr$, $K = (r^2+a^2)\omega - a m$, and 
${}_s \lambda_{lm}$ is the spheroidal eigenvalue.  The relevant physical solutions, labelled ${}_{-2} R^\textrm{in}_{\ell m \omega}(r)$, satisfy the boundary condition demanding that they consist of purely ingoing radiation at the event horizon $r=r_+:=GM+\sqrt{(GM)^2-a^2}$,
\begin{equation}\label{Eq:TeukBC}
	{}_{-2} R^\textrm{in}_{\ell m \omega}=B^\textrm{trans}_{{\ell}{m}{\omega}} \Delta^2e^{-i\tilde{\omega} r_*} \;\;\textrm{as}\;\; r\rightarrow r_+,
\end{equation}
where $r_*$ is the tortoise coordinate and $\tilde\omega=\omega-m a/(2G M r_+)$.  This fixes the asymptotic behavior at radial infinity to be of the form
\begin{align}
\label{Eq:InfFormTeuk}
	{}_{-2} R^\textrm{in}_{\ell m \omega}&=B^\textrm{inc}_{{\ell}{m}{\omega}} r^{-1}e^{-i\omega r_*}+B^\textrm{ref}_{{\ell}{m}{\omega}} r^3 e^{i\omega r_*}  \\& \;\;\textrm{as}\;\; r\rightarrow \infty,\nnm  
\end{align}
where $B^\textrm{inc}_{{\ell}{m}{\omega}}$ and $B^\textrm{ref}_{{\ell}{m}{\omega}}$\footnote{Note that the above form is identical to the one used earlier in Eq.~(\ref{intruth}) except that the parity dependent factors have been absorbed into $B_{lm\omega}^{\mr{inc}}$. As a result, the expression for the scattering phase now contains a parity-dependent factor [see Eq.~(\ref{dolanphase})]. This maybe a more convenient convention for BHPT but the former is more transparent for matching with the effective theory.} are the coefficients of the incident and reflected waves.  The ratio $B^\textrm{ref}_{{\ell}{m}{\omega}}/B^\textrm{inc}_{{\ell}{m}{\omega}}$ is completely determined by demanding that ${}_{-2} R^\textrm{in}_{\ell m \omega}(r)$ solve the radial Teukolsky equation (\ref{Eq:RadTeuk}) with the boundary condition (\ref{Eq:TeukBC}).  We refer the reader to Ref.~\cite{Sasaki:2003xr} for details of a procedure to produce the expansion of this ratio in powers of $GM\omega$.  Finally, the relevant scattering phase shifts and transmission factor $\eta^P_{lm}e^{2i\delta_{lm\omega}^P}$ (equivalent to $\mathcal{C}^{lm}_{P,\mr{out}}/\mathcal{C}^{lm}_{P,\mr{in}}$ from the effective theory above) for waves of parity $P=\pm 1$ are given, e.g.\ as in Eq.~(30) of Ref.~\cite{Dolan:2008kf}, by
\begin{alignat}{3}
\label{dolanphase}
\nnm\eta_{lm}^P \exp(2i\delta_{lm\omega}^P)&=(-1)^{l+1}\left(\frac{\textrm{Re}(C)+12iGM\omega P}{16\omega^4}\right)\\ &\times\frac{B^\textrm{ref}_{{\ell}{m}{\omega}}}{B^\textrm{inc}_{{\ell}{m}{\omega}}}, 
\end{alignat}
with $[\textrm{Re}(C)]^2$ as given in Eq.~(31) of Ref.~\cite{Dolan:2008kf}.

 This yields the complete expression for the degree of absorption for $l=2$ and $l=3$ spheroidal modes up to $\mathcal{O}(\epsilon^7)$ to be
 \begin{widetext}
 \begin{alignat}{3}
 \label{tru2}
1-|\eta_{2m}^{P=\pm}| =& -\epsilon^5 m\bigg[\frac{2 \mathcal{A}_{0}^1}{5}-(m^2-4)\frac{2}{15}\mathcal{A}_{0}^3\bigg](1+2\epsilon \pi)\\& \nnm +\epsilon^6\bigg\{\frac{4 \mathcal{A}^0_{1} }{5}-(m^2-4)\bigg[\frac{2 \mathcal{A}^2_{1}}{15}+\frac{2 \mathcal{A}^4_{1}}{15}(m^2-1)\bigg]\bigg\}  + \mathcal{O}(\epsilon^7), \\
1-|\eta_{3m}^{P=\pm}| =& \mathcal{O}(\epsilon^7),
\label{tru3}
 \end{alignat}
 where 
 \begin{alignat}{3}
\label{vals}
&\mathcal{A}^1_0 = \frac{16\chi}{45}(1+3\chi^2), ~ \mathcal{A}_{0}^3 = -\frac{4\chi^3}{3},\\& \nnm \mathcal{A}_{1}^0 = \frac{16}{405}(9+9\kappa+97\chi^2+117\kappa\chi^2-6\chi^4+54\sigma\chi^4+36\chi B_2 + 108\chi^3B_2), \\& \nnm \mathcal{A}_{1}^2=-\frac{8}{135}(115\chi^2+135\kappa\chi^2+5\chi^4+90\kappa\chi^4-24\chi B_1+ 18 \chi^3 B_1+48\chi B_2 + 144\chi^3 B_2), \\& \nnm \mathcal{A}_{1}^4  = \frac{8}{135}(20\chi^4+45\kappa\chi^4-24\chi B_1+18\chi^3 B_1+12\chi B_2+36 \chi^3 B_2),
\end{alignat}
 \end{widetext}
 and we are using the notation $B_m = \mr{Im}[\mr{PolyGamma}(0,3+im\chi/\kappa)]$, which is odd in $\chi$, and $\kappa=\sqrt{1-\chi^2}$. Note that the degree of absorption obtained from solving the Teukolsky equation in Eqs.~(\ref{tru2}), (\ref{tru3}) has almost exactly the same form as that obtained from the effective theory given in Eq.~(\ref{weftabs}), with the only difference being the factor of $(1+2\epsilon \pi)$ next to the leading-order $\epsilon^5$ result for $l=2$ mode. This factor is missing in the degree of absorption derived in the effective theory in Eq.~(\ref{weftabs}) due to us neglecting the leading-order nonlinear interaction between the GW and the gravitational field of the particle. In principle, this can be also obtained from the effective worldline theory by including the nonlinearities and regulating any resulting divergences. However, including the leading-order nonlinearities in this classical setup which we are using is a complicated task, and not very illuminating. It is easier instead to just replace the factor with $1$ on the Teukolsky side by tracing its origins to the nonlinearities neglected in the scattering problem in the effective picture. We establish this explicitly for the simpler case of a scalar field scattering of a Schwarzschild BH by including the leading-order nonlinearities on the effective theory side in Appendix.~\ref{regcorr}. There are essentially two physical processes involved in the effective theory picture, both arising from the interaction of the external wave (gravitational or otherwise) with the stationary gravitational field sourced by the particle. The first is that the value of the tidal field of the wave at the origin (location of the particle) is modified, as shown for a scalar wave in Eq.~(\ref{regch}) by a factor of $(1+\pi\epsilon)$. This in turn modifies the strength of the tidally induced multipole moment subsequently affecting the irregular (output) part of the wave and the degree of absorption. Additionally, the wave is modified in its journey away (to) the particle to (from) infinity again due to scattering off the particle's gravitational field. This factor modifies the form of the wave asymptotically far away from the particle, as shown again for a scalar wave in Eq.~(\ref{sasymp}) in such a way that the degree of absorption is further multiplied by a factor of $(1+\pi\epsilon)$. This result has also been derived for the case of GW amplitudes sourced by arbitrary multipole moments in Refs.~\cite{Goldberger:2009qd,Porto:2016pyg} and is seen to modify the radiated power by the square of that factor. Together, these two effects modify the leading-order degree of absorption by a factor $(1+\epsilon\pi)^2\approx (1+2\epsilon \pi) + \mathcal{O}(\epsilon^2)$ which is seen in Eq.~(\ref{tru2}). We have in fact verified the presence of this factor multiplying the leading-order degree of absorption for all different $l$ modes (for which we had the solution) for scalar, and tensor (gravitational) fields, specifically $l=0,~1,~2,~3$ modes for the scalar field and $l=2,~3$ for gravitational field. Thus, to match the part of the true degree of absorption which corresponds to the flat space-time scattering in the effective picture, it is sufficient to simply replace the $(1+2\pi\epsilon)$ factor for $1$ from Eq.~(\ref{tru2}).
 
 Thus, dropping the $(1+2\epsilon\pi)$ factor from Eq.~(\ref{tru2}), and then comparing it with Eq.~(\ref{weftabs}), we fix the unknown coefficients to be
\begin{widetext}
\begin{alignat}{3}
\label{fixed}
&f^1_{\mce/\mcb,0} = \frac{16\chi}{45}(1+3\chi^2), ~ f^3_{\mce/\mcb,0} = -\frac{4\chi^3}{3},\\& \nnm f^0_{\mce/\mcb,1} = \frac{16}{405}(9+9\kappa+97\chi^2+117\kappa\chi^2-6\chi^4+54\kappa\chi^4+36\chi B_2 + 108\chi^3B_2), \\& \nnm f^2_{\mce/\mcb,1}=-\frac{8}{135}(115\chi^2+135\kappa\chi^2+5\chi^4+90\kappa\chi^4-24\chi B_1+ 18 \chi^3 B_1+48\chi B_2 + 144\chi^3 B_2), \\& \nnm f^4_{\mce/\mcb,1} = \frac{8}{135}(20\chi^4+45\kappa\chi^4-24\chi B_1+18\chi^3 B_1+12\chi B_2+36 \chi^3 B_2), \\&
 \nnm g_{\mcb,0}^{i} = h^i_{\mce,0} = \frac{2}{3}\chi f_{\mce,0}, \\& g_{\mce,0}^{i} = h^i_{\mcb,0} = -\frac{2}{3}\chi f_{\mcb,0}, \label{const2}
\end{alignat}
\end{widetext}
where we have also restated the constraints obtained by imposing spheroidal separability from Eq.~(\ref{const}). Having fixed the (dissipative) response coefficients, we can now compute the (dissipative part of the) induced multipole moments in any setup. In particular, we can now consider the effect of induced tides in a binary system with two spinning BHs in the inspiral phase. Our focus is on computing the change in mass and angular momentum due to tidal effects in the worldline effective theory (and horizon fluxes in the real setup). In the next section, we derive general formulae for evolution equations of mass and spin and then specialize to the case of parallel-spin--quasi-circular orbits, which can then be compared with earlier results available in literature.
\section{General expressions for evolution equations of mass and spin}
\label{dmjdt}
In this section, we derive general formulae for computing the evolution of  of mass and spin from the equations of motion. Then, we proceed to compute them explicitly using the now fixed response coefficients for the special case of parallel-spin--quasi-circular inspiral to relative 1.5PN order. We conclude this section by comparing these results with those obtained earlier in Refs.~\cite{Alvi:2001mx, Chatziioannou:2016kem, Tagoshi:1997jy}.

We can derive the formula for mass and spin evolution from the equations of motion for momentum and spin angular momentum respectively. For mass, we start with the equation of motion for momentum, i.e., 
\begin{alignat}{3}
\nnm \frac{Dp^{\mu}}{D\tau} &= -\frac{1}{2}R_{\mu\nu\rho\sigma}u^{\nu}S^{\rho\sigma} - \frac{1}{6}J^{\lambda\nu\rho\sigma}\nabla_{\mu}R_{\lambda\nu\rho\sigma} \\& - J^{\tau\lambda\nu\rho\sigma}\frac{1}{12}\nabla_{\mu}\nabla_{\tau}R_{\lambda\nu\rho\sigma},
\end{alignat}
where we substitute the expressions for $J^{\mu\nu\rho\sigma}$ and $J^{\lambda\mu\nu\rho\sigma}$ given in Eqs.~(\ref{4quad}), (\ref{4oct}), yielding
\begin{alignat}{3}
\frac{Dp^{\mu}}{D\tau} \nnm &= -\frac{1}{2}R_{\mu\nu\rho\sigma}u^{\nu}S^{\rho\sigma} - \frac{1}{2}Q^{\rho\sigma}\nabla^{\mu}\mce_{\rho\sigma}- \frac{1}{2}Q_\mcb^{\rho\sigma}\nabla^{\mu}\mcb_{\rho\sigma} \\& - \frac{1}{2}Q_\mce^{\rho\sigma\lambda}\nabla^{\mu}\mce_{\rho\sigma\lambda}- \frac{1}{2}Q_\mcb^{\rho\sigma\lambda}\nabla^{\mu}\mcb_{\rho\sigma\lambda}.
\end{alignat}
Now, we define in the effective worldline theory the mass $m$ simply as $\sqrt{-p^2}$. Then we have 
\begin{alignat}{3}
&p_{\mu}\frac{Dp^{\mu}}{D\tau} = -m \frac{dm}{d\tau} \nnm \\ & \nnm \implies  \frac{dm}{d\tau} = \frac{1}{2m}p^{\mu}u^{\nu}R_{\mu\nu\rho\sigma}S^{\rho\sigma} + \frac{1}{2m}Q_\mce^{\rho\sigma}(p\cdot\nabla)\mce_{\rho\sigma} \\& \nnm +\frac{1}{2m}Q_\mcb^{\rho\sigma}(p\cdot\nabla)\mcb_{\rho\sigma} +\frac{1}{2m}Q_\mce^{\rho\sigma\lambda}(p\cdot\nabla)\mce_{\rho\sigma\lambda}\\&+\frac{1}{2m}Q_\mcb^{\rho\sigma\lambda}(p\cdot\nabla)\mcb_{\rho\sigma\lambda},
\end{alignat}
and the first term can be shown to vanish using the relation between $p^{\mu}$ and $u^{\mu}$ if we neglect terms cubic or higher powers in curvature. They are not relevant to the relative 1.5PN order (in horizon fluxes) we are interested in this work. Similarly, we can substitute $p^{\mu}=m u^{\mu}$ in the terms with multipole moments to the order of interest to get 
\begin{alignat}{3}
\frac{dm}{d\tau} = \frac{1}{2}Q_{\mce}^{\rho\sigma}\dot{\mce}_{\rho\sigma} + \frac{1}{2}Q_{\mce}^{\rho\sigma\lambda}\dot{\mce}_{\rho\sigma\lambda} + (\mce\leftrightarrow \mcb)
\label{massloss}
\end{alignat}
which gives us the general formula for mass evolution valid to relative 1.5PN order. We expect this to match with the horizon energy flux up to any total time derivatives of functions of tidal fields, which should vanish for scattering events or for quasi periodic processes (like parallel-spin--quasi-circular inspiral which we shall consider shortly). The quadrupolar contribution to mass-change matches with that given in Ref.~\cite{Goldberger:2020fot} if we identify our quadrupole tensors as twice of theirs. This is because they choose a different normalization in the tidal coupling terms. \footnote{ In Ref.~\cite{Goldberger:2020fot}, the quadrupole tidal coupling terms in the action are $Q_\mce^{\mu\nu}\mce_{\mu\nu} + (\mce\leftrightarrow \mcb)$. Thus, before comparing the expressions in this work with that in Ref.~\cite{Goldberger:2020fot}. One must first transform as $Q_{\mce(\mcb)}^{\mu\nu} \rightarrow 2 Q_{\mce(\mcb)}^{\mu\nu}$\label{impfut}}

We can similarly derive the equation for spin evolution from the equation of motion for spin angular momentum, i.e.,
\begin{widetext}
\begin{alignat}{3}
\nnm \frac{DS^{\mu\nu}}{D\tau} &= 2 p^{[\mu}u^{\nu]} + \frac{4}{3}R^{[\mu}{}_{\tau\rho\sigma}J^{\nu\tau\rho\sigma} + \frac{2}{3}\nabla^{\lambda}R^{[\mu}{}_{\tau\rho\sigma}J_{\lambda}{}^{\nu\tau\rho\sigma}+\frac{1}{6}\nabla^{[\mu}R_{\lambda\tau\rho\sigma}J^{\nu]\lambda\tau\rho\sigma}\nnm \\
J\frac{DJ}{d\tau}=\frac{1}{2}S_{\mu\nu}\frac{DS^{\mu\nu}}{D\tau} &= \frac{2}{3}R^{\mu}{}_{\tau\rho\sigma}J^{\nu\tau\rho\sigma}S_{\mu\nu} + \frac{1}{3}\nabla^{\lambda}R^{\mu}{}_{\tau\rho\sigma}J_{\lambda}{}^{\nu\tau\rho\sigma}S_{\mu\nu}+\frac{1}{12}\nabla^{\mu}R_{\lambda\tau\rho\sigma}J^{\nu\lambda\tau\rho\sigma}S_{\mu\nu} \nnm \\
\frac{DJ}{d\tau} &= \frac{1}{J}Q_{\mce}^{\mu\nu} \mce_{\mu}{}^{\rho}S_{\rho\nu} +\frac{3}{2J}O_{\mce}^{\mu\nu\lambda}\mce_{\mu\lambda}{}^{\rho}S_{\nu\rho} + (\mce
\leftrightarrow \mcb),
\label{angloss}
\end{alignat}
\end{widetext}
where we have defined $J^2=(1/2)S^{\mu\nu}S_{\mu\nu}$ as the magnitude of spin angular momentum of the BH. Again, the quadrupolar contribution to evolution of spin($dJ/dt$) is identical to that in Ref.~\cite{Goldberger:2020fot} once the multipole moments are properly identified (see footnote.~\ref{impfut}).

Now, substituting the ans{\"a}tze for the multipole moments from Eqs.~(\ref{moctet}) into the expressions for the evolution of mass and spin in Eqs.~(\ref{massloss}), (\ref{angloss}). We get 
\begin{widetext}
\begin{alignat}{3}
\label{expdm}
\nnm \frac{dm^{*}}{d\tau} & = \frac{m}{2}(Gm)^4 \{f^1_{\mce,0} (\dot{\mathcal{E}}^{\mu\nu}\mathcal{E}_{\mu}{}^{\rho}\hS_{\nu\rho})  + f^3_{\mce,0}(\dot{\mathcal{E}}_{\mu}{}^{\rho}\mathcal{E}_{\nu}{}^{\sigma}\hs^{\mu}\hs^{\nu}\hS_{\rho\sigma})\\&+ \nnm (Gm)[f_{\mce,1}^{0} (\dot{\mathcal{E}}^{\mu\nu}\dot{\mathcal{E}}_{\mu\nu})+f_{\mce,1}^{2}(\dot{\mathcal{E}}_{\mu}{}^{\rho}\dot{\mathcal{E}}_{\nu\rho}\hs^{\mu}\hs^{\nu}) + f_{\mce,1}^{4} (\dot{\mathcal{E}}_{\mu\nu}\dot{\mathcal{E}}_{\rho\sigma}\hs^{\mu}\hs^{\nu}\hs^{\rho}\hs^{\sigma})\\&+\xi\frac{2}{3}\chi f_{\mce,0}^1(\mathcal{B}_{\mu\rho\sigma}\dot{\mathcal{E}}^{\nu\rho} - \dot{\mathcal{B}}_{\mu\rho\sigma}\mathcal{E}^{\nu\rho})\hs^{\mu}\hS_{\nu}{}^{\sigma}+\xi\frac{2}{3}\chi f_{\mce,0}^3(\mathcal{B}_{\nu\rho\lambda}\dot{\mathcal{\mce}}_{\mu}{}^{\sigma}-\dot{\mathcal{B}}_{\nu\rho\lambda}\mathcal{E}_{\mu}{}^{\sigma})\hs^{\mu}\hs^{\nu}\hs^{\rho}\hS_{\sigma}{}^{\lambda}]\} \nnm
 \\& + (\mathcal{\mce}\leftrightarrow \mathcal{\mcb}, ~ \xi \rightarrow -\xi), \\
 \label{expdJ}
\nnm \frac{dJ^{*}}{d\tau} &= \frac{M}{2}(Gm)^4\{-2 f_{\mce,0}^1(\mce_{ij}\mce^{ij})+(3 f_{\mce,0}^1-f_{\mce,0}^3)(\mce_{i}{}^{k}\mce_{jk}\hs^{i}\hs^{j})+f_{\mce,0}^3 (\mce_{ij}\hs^{i}\hs^{j})^2
\\& \nnm - (Gm)[f_{\mce,1}^{0} (\dot{\mathcal{E}}^{\mu\nu}\mathcal{E}_{\mu}{}^{\rho}\hS_{\nu\rho})+f_{\mce,1}^2 (\dot{\mathcal{E}}_{\mu}{}^{\rho}\mathcal{E}_{\nu}{}^{\sigma}\hs^{\mu}\hs^{\nu}\hS_{\rho\sigma}) \\& + \xi\frac{2}{3}\chi f_{\mce,0}^1 (\mathcal{B}_{\mu\sigma\lambda}\mathcal{E}^{\nu\rho}\hs^{\mu}\hS_{\nu}{}^{\sigma}\hS_{\rho}{}^{\lambda}) + \xi \frac{2}{3} \chi f_{\mce,0}^1 (\mathcal{B}_{\mu\rho\lambda}\mce^{\nu\rho}\hs^{\mu}\hS_{\nu}{}^{\sigma}\hS_{\sigma}{}^{\lambda}) + \xi\frac{2}{3}\chi f_{\mce,0}^3 (\mcb_{\nu\rho\tau}\mce_{\mu}{}^{\sigma}\hs^{\mu}\hs^{\nu}\hs^{\rho}\hS_{\sigma}{}^{\lambda}\hS_{\lambda}{}^{\tau})]\}  \nnm \\& + (\mce\leftrightarrow\mcb,~\xi\rightarrow -\xi) ,
\end{alignat}
where dots represent covariant derivatives w.r.t proper time, $\xi=1$, and we have imposed the constraints obtained by demanding spheroidal separability from Eq.~(\ref{const}). Additionally, we have
\begin{alignat}{3}
\label{mshift}
m^{*} &=  m - \frac{m}{4}(Gm)^4\{f_{\mce,0}^0 (\mce^{\mu\nu}\mce_{\mu\nu}) + f_{\mce,0}^2 (\mce_{\mu}{}^{\rho}\mce_{\nu\rho}^{\sigma}\hs^{\mu}\hs^{\nu})+f_{\mce,0}^4 (\mce_{\mu\nu}\mce_{\rho\sigma}\hs^{\mu}\hs^{\nu}\hs^{\rho}\hs^{\sigma}) \nnm \\& + (Gm)[\xi\frac{2}{3}\chi f_{\mce,0}^0(\mcb_{\mu\nu\rho}\mce^{\nu\rho}\hs^{\mu}) + \xi\frac{2}{3}\chi f_{\mce,0}^2(\mcb_{\nu\rho\sigma}\mce_{\mu}{}^{\sigma}\hs^{\mu}\hs^{\nu}\hs^{\rho})+\xi\frac{2}{3}\chi f_{\mce,0}^4(\mcb_{\rho\sigma\lambda}\mce_{\mu\nu}\hs^{\mu}\hs^{\nu}\hs^{\rho}\hs^{\sigma}\hs^{\lambda}) ]\}\nnm \\& + (\mce \leftrightarrow \mcb,~\xi\rightarrow -\xi),\\
\label{jshift}
J^{*} & = J - \frac{M}{4}(GM)^5\{f_{\mce,1}^1 [(\mce^{\mu\nu}\mce^{\rho\sigma}\hS_{\mu\rho}\hS_{\nu\sigma})-(\mce^{\mu\nu}\mce_{\mu}{}^{\rho}\hS_{\nu}{}^{\sigma}\hS_{\rho\sigma})]- f_{\mce,1}^3(\mce_{\mu}{}^{\rho}\mce_{\nu}{}^{\sigma}\hs^{\mu}\hs^{\nu}\hS_{\rho}{}^{\lambda}\hS_{\sigma\lambda})	\},
\end{alignat}
\end{widetext}
and we see that all the tidal coefficients that do not enter the degree of absorption in the RHS of Eq.~(\ref{weftabs}) only shift the definition of mass and angular-momentum by quadratic functions of fields (i.e., they only contribute terms that are total-time derivatives to $dm/dt$ and $dJ/dt$). Whereas the ones that do show up in degree of absorption contribute terms that cannot be absorbed as such in total time-derivatives. Thus, for any quasi periodic setup (like for parallel-spin--quasi-circular orbits) or in a scattering set up where the two particles are infinitely far in the distant past or future, the average or total change in mass/spin respectively is determined entirely by the coefficients that contribute to dissipation, which we have already fixed through comparison with the degree of absorption obtained in the real theory by solving the Teukolsky equation in Sec.~\ref{match}. This is since the total-time derivative terms either vanish (when the particles are far away) or cancel (in a periodic setup). In deriving the above result, we have also used the fact that the covariant time-derivative of spin tensor and vector (which enter the ans{\"a}tze) vanishes up to the relative order to which we have expanded the expressions for $dm/dt$ and $dJ/dt$ (i.e., up to relative 1.5PN). 

The expressions for mass and spin evolution in Eqs.~\eqref{expdm}, \eqref{expdJ} can be compared with those in Refs.~\cite{Goldberger:2020fot, Chatziioannou:2016kem} by substituting the response coefficients from Eq.~(\ref{fixed}), and we indeed find that our expressions are identical at leading order in $GM$ (i.e., the $(GM)^4$ part), but differs from Ref.~\cite{Chatziioannou:2016kem} at next order in $GM$. A crucial difference in our expressions when compared with those in Ref.~\cite{Chatziioannou:2016kem} is that octupolar tidal fields do not enter their expressions at all. Another interesting difference is that there are $\pi^2$-containing coefficients in their next-to-leading order expression for $dJ/dt$ (and subsequently in $dm/dt$), whereas $\pi^2$ does not enter any of our tidal-response coefficients. However, we will see later in Sec.~\ref{victory} that our expression for mass and spin evolution is consistent with Ref.~\cite{Tagoshi:1997jy} for the special case of a test-body in a circular orbit around a Kerr BH to relative 1.5PN order, unlike Ref.~\cite{Chatziioannou:2016kem}.

In the next section, we specialize to the parallel-spin--quasi-circular setting to compute the expression mass and spin evolution during inspiral up to 1.5PN relative to the same at leading order (4PN w.r.t leading-order flux to infinity), and compare with earlier works that produced expressions for the same. 

\subsection{Results for the special case of binaries with parallel spins in and circular orbits}

\begin{figure}[h]
\begin{center}
\vspace{0.3truecm}
\includegraphics[width=0.7\linewidth]{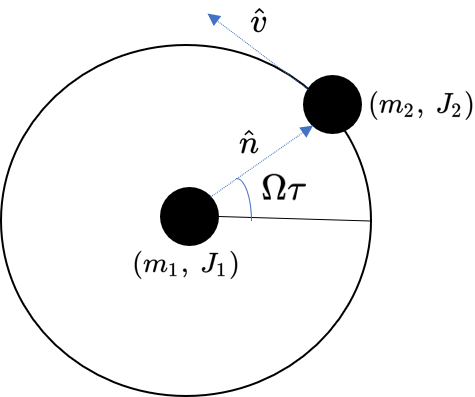}
\caption{A graphic illustrating a parallel-spin--quasi-circular binary with two BHs with masses $m_1$ and $m_2$, and spins $J_1$ and $J_2$. The spin vectors are orthogonal to the orbital plane. The image is drawn in the comoving frame of BH with mass $m_1$, with $\hat{n}$ being the unit-vector pointing towards the other BH ($m_2,~J_2$) and $\hat{v}$ being the unit-vector along the other BH's velocity. $\Omega$ is the angular velocity of the other BH and hence of the tidal fields in this frame. $\tau$ is the proper time of mass $m_1$.}
\label{fig1}
\vspace{0.3truecm}
\end{center}
\end{figure}
Now, we consider a system of two BHs with initial masses $m_1$ and $m_2$ and spin parameters $\chi_1$ and $\chi_2$. Their spins are parallel and orthogonal to the orbital plane and they are in a quasi-circular orbit. We can compute the rate of change of (initial) mass $m_1$ and spin $J_1=Gm_1^2\chi_1$ averaged over one orbit using the equations Eqs.~(\ref{massloss}),(\ref{angloss}). In this section, we will refer to the BH with initial mass $m_1$ and $J_1$ as the primary BH, and the other BH as secondary here onwards. The tidal fields are sourced by the other BH of mass $m_{2}$ with spin parameter $\chi_{2}$, although they are affected by nonlinear interaction with the fields due to the primary BH ($m_1$, $J_1$). This is most conveniently done when the tidal fields are computed in a locally flat rest frame of the primary BH (see Fig.~\ref{fig1}), since then the covariant derivatives of the field w.r.t proper time can be treated simply as ordinary time derivatives . This has already been done for the quadrupolar fields $\mce^{\mu\nu}$ and $\mcb^{\mu\nu}$ in Ref.~\cite{Chatziioannou:2016kem}, and we simply borrow the expressions from there. Rewriting the expressions here here, we have 
\begin{widetext}
\begin{alignat}{3}
\nnm\frac{1}{2}(\mathcal{E}_{11}+\mathcal{E}_{22})&=-\frac{m_2}{2r^3}\bigg[1+\frac{X_1}{2} V^2-6X_{2}\chi_{2}V^3+\mathcal{O}(V^4)\bigg],\\
\nnm\frac{1}{2}(\mathcal{E}_{11}-\mathcal{E}_{22})&=-\frac{3m_{2}}{2r^3}\bigg[1+\frac{X_1-4}{2}V^2-2X_{2}\chi_{2}V^3+\mathcal{O}(V^4)\bigg]\cos(2\Omega \tau), \\
\mathcal{E}_{12} &=-\frac{3m_{2}}{2r^3}\bigg[1+\frac{X_1 -4}{2}V^2-2X_{2}\chi_{2}V^3+\mathcal{O}(V^4)\bigg]\sin(2\Omega \tau), \label{equad}\\
\nnm\mathcal{B}_{13}&=\frac{-3m_{2}}{r^3}V(1-X_{2}\chi_{2}V)\cos(\Omega \tau) + \mathcal{O}(V^3), \\
\mathcal{B}_{23}&=\frac{-3m_{2}}{r^3}V(1-X_{2}\chi_{2}V)\sin(\Omega \tau) + \mathcal{O}(V^3), \label{bquad}
\end{alignat}
where $\Omega$ is the angular velocity of the tidal field in the primary BH frame ($m_1$) given by
\begin{alignat}{3}
\Omega &=\sqrt{\frac{M}{r^3}}\bigg[1-\frac{1}{2}(3+\eta)V^2-\frac{1}{2}\bar{\chi}V^3+\mathcal{O}(V^4)\bigg],\\
\bar{\chi} &= X_1(1+X_1)\chi+3\eta\chi_2,
\end{alignat}
\end{widetext}
and we are using the notation : $M=m_1+m_{2}$ and $X_1=m_1/M$, $X_{2}=m_{2}/M$ are the mass-fractions. $\eta=X_1X_2$ is the symmetric mass ratio, $r$ is the orbital separation in harmonic coordinates and $V=\sqrt{\frac{M}{r}}$. Here, we are working in units with $G=c=1$, as was done in Ref.~\cite{Chatziioannou:2016kem} so that we can easily compare our results although we have changed the notation quite a bit. Additionally, in Ref.~\cite{Chatziioannou:2016kem}, a sign factor $\epsilon=\pm 1$ was used in front of the expression for $\Omega$, to denote whether the secondary BH was spin aligned or antialigned w.r.t the orbital angular momentum. However, we simply let the spin parameter(s) $\chi_{1,2}$ range over [-1,1] (instead of [0,1]) and always fix the orbital angular momentum to be aligned along the positive z-axis without loss of generality.

The octupolar fields were not derived in Ref.~\cite{Chatziioannou:2016kem} since they were not relevant in their expression for the mass or spin evolution. For us, the octupolar fields do contribute to the expressions for mass and spin evolution as seen from Eqs.~(\ref{expdm}),~(\ref{expdJ}). Fortunately, they (octupolar fields) are only relevant at leading order and thus can be easily computed from the test-body limit at leading post Newtonian order where the secondary BH (as test mass) is orbiting the primary BH in the limit $m_2\ll m_1$. The formula for leading-order fields does not change from this for generic mass-ratio. We get the expressions 
\begin{alignat}{3}
\mathcal{E}_{ijk} & = -15 \frac{m_{2}}{r^4}\hat{n}_{\langle i} \hat{n}_j \hat{n}_{k\rangle}, \label{eoct}\\
\mathcal{B}_{ijk} & = 30\frac{m_{2}}{r^4}V\epsilon_{\langle i}{}^{mn}\hat{n}_{j}\hat{n}_{k\rangle}\hat{n}_{m}\hat{v}_{n},
\label{boct}\\
\hat{n} &= (\hat{n}_1, \hat{n}_2, \hat{n}_3) = (\cos(\Omega \tau), \sin(\Omega \tau), 0),\quad \nnm \\\hat{v} & = (\hat{v}_1, \hat{v}_2, \hat{v}_3) = (-\sin(\Omega \tau), \cos(\Omega \tau), 0) \nnm  
\end{alignat} 
where $n_{i}$ is the normal vector directed from the primary BH to the secondary BH in harmonic coordinates, and $\hat{v} =  \dot{\hat{n}}$ is the relative velocity vector of the secondary BH w.r.t primary BH. 

Now, substituting these tidal fields from Eqs.~(\ref{equad}), (\ref{bquad}), (\ref{eoct}), and (\ref{boct}) into the formulas for mass and spin evolution in Eqs.~(\ref{expdm}),~(\ref{expdJ}), with the fixed coefficients listed in Eqs.~(\ref{fixed}),~(\ref{const2}), we get the orbit-averaged results 
\begin{widetext}
\begin{alignat}{3}
\label{mlcircgV}\Big\langle\frac{dm_1}{d\tau}\Big\rangle &= \Omega(\Omega_\mr{H}-\Omega)C_{V}, \\ \label{jlcircgV}\Big\langle\frac{dJ_1}{d\tau}\Big\rangle &= (\Omega_\mr{H}-\Omega)C_{V},  \\
\label{alcircgV}
\Big\langle\frac{dA_1}{d\tau}\Big\rangle &= \frac{(dm_1- \Omega_\mr{H} dJ_1)}{d\tau}\frac{8 \pi}{\kappa} = \frac{-8\pi}{\kappa}(\Omega_H-\Omega)^2C_V ,
\end{alignat}
where $\Omega_\mr{H}=\chi_1/[2m_1(1+\kappa_1)]$, is the horizon angular velocity of the primary BH and $A_1=8\pi m_1^2(1+\kappa)$, is the horizon-surface area and 
\begin{alignat}{3}
\nnm C_V= &-\frac{16}{5}m_1^2 X_1^2 \eta^2 (1+\kappa_1)V^{12} \Big\{1+3\chi_1^2 +V^2\big(-3+X_1-\frac{51}{4}\chi_1^2+3X_1\chi_1^2\big)\\& +V^3\Big[\frac{-X_1\chi_1}{3}(64+60\kappa_1+33\chi_1^2+36\kappa_1\chi_1^2)-\frac{3}{2}X_{2}(4+7\chi_1^2)\chi_{2}-8X_1(1+\chi_1^2)B_2(\chi_1)\Big]\Big\}.
\end{alignat}
\end{widetext}
The contributions to the definition of mass and angular momentum due to the conservative tidal coefficients as seen in Eqs.~(\ref{mshift}), (\ref{jshift}) were removed upon averaging over an orbit. They will also not contribute in a scattering scenario provided we can set the tidal fields to zero along the worldlines of the particle asymptotically (in the distant past and future).

However, the results in Eqs.~(\ref{mlcircgV}), (\ref{jlcircgV}), and (\ref{alcircgV}) are written in terms of gauge-dependent quantities, namely $r$ (which enters through $V$) and is the separation between the two bodies in harmonic coordinates. Further more, it is written in the frame of the primary BH as opposed to the more convenient PN barycentric frame (which coincides with the primary BH frame in the test-body limit for the secondary BH). Thus, before comparison, we convert the results to the PN barycentric frame using the relations 
\begin{alignat}{3}
 V &= x\Big[1+\frac{1}{6}(3-\eta)x^2+\frac{1}{6}\tilde{\chi}x^3+\mathcal{O}(x^4)\Big], \label{V2x} \\
 t &= \tau\Big[1+\frac{1}{2}(2X_1+3X_{2})X_{2}x^2+\mathcal{O}(x^4)\Big], \label{timrel}
\end{alignat} 
where $x = (M \omega_{\mr{orb}})^{1/3}$ with $\omega_{\mr{orb}}$ is the orbital angular velocity, which is gauge invariant PN expansion parameter and $\tilde{\chi}=(2 X_1^2+3\eta)\chi_1+(3\eta+2X_2^2)\chi_2$. $t$ is the PN barycentric time and its relation to the proper time of the primary BH $\tau$, is given in Eq.~(\ref{timrel}). These expressions have been taken from Eqs.(39) and (40) in Ref.~\cite{Chatziioannou:2016kem}. Then, the expressions in Eqs.~(\ref{mlcircgV}), (\ref{jlcircgV}), and (\ref{alcircgV}) can be rewritten in the PN barycentric frame as an expansion in the gauge invariant parameter $x$ as 
\begin{widetext}
\begin{alignat}{3}
\Big\langle\frac{dm_1}{dt}\Big\rangle &= \Omega(\Omega_\mr{H}-\Omega)C_{x} , \label{mlcircg}\\ \Big\langle\frac{dJ_1}{dt}\Big\rangle &= (\Omega_\mr{H}-\Omega)C_{x} \label{jlcircg},  \\
\Big\langle\frac{dA}{dt}\Big\rangle &= \frac{(dm_1- \Omega_\mr{H} dJ_1)}{dt}\frac{8 \pi}{\kappa} = \frac{-8\pi}{\kappa}(\Omega_H-\Omega)^2C_x ,
\end{alignat}
where
\begin{alignat}{3}
C_x= &-\frac{16}{5}M^2 X_1^2 \eta^2 (1+\kappa_1)x^{12} \Big(1+3\chi_1^2 + \frac{1}{4}[3(2+\chi_1^2)+2X_1(1+3\chi_1^2)(2+3X_1)]x^2 \\&+x^3\Big\{\frac{1}{2}(-4+3\chi_1^2)\chi_2-2 X_1(1+3\chi_1^2)(X_1(\chi_1+\chi_{2})+4 B_2(\chi_1))+X_1\Big[-\frac{2}{3}(23+30\kappa_1)\chi_1+(7-12\kappa_1)\chi_1^3+4\chi_{2}\nnm \\& +\frac{9\chi_1^2\chi_{2}}{2}\Big]\Big\}\Big), \nnm
\end{alignat}
\end{widetext}
which can now be conveniently compared with the expressions for the same quantities given in  Eq.~(45) of Ref.~\cite{Chatziioannou:2016kem}. We find that our expression for $C_x$ for generic-mass ratios  is consistent with the result \cite{Chatziioannou:2016kem} to NLO (to $x^{14}$) but not at NNLO (at $x^{15}$). An important visible difference is that we have no $\pi^2$-containing terms at NNLO. However, as we will see in the next subsection, it is consistent in the test body limit with earlier results computed by solving the Teukolsky equation for the curvature perturbation sourced by a test-body moving in a circular orbit around a spinning BH up to relative 1.5PN order ($x^{15}$).
\subsection{The test body limit for circular orbits with parallel spins}
\label{victory}
In the special case where the other BH  with mass $m_2$ becomes a test particle, we only evaluate the quantities to leading order in $X_{2}$, mass ratio of other particle. This is equivalent to simply setting $X_1=1$, $X_2\rightarrow 0$ and thus $M=m_1\rightarrow \infty$ such that $M X_2=m_{2}$ remains constant.  Then the result simplifies to
\begin{widetext}
\begin{alignat}{3}
\frac{dm_1}{dt} =  & \mathcal{F}_{\infty}\bigg\{-\frac{x^5}{4}(\chi_1+3\chi_1^3)-x^7(\chi_1+\frac{33}{16}\chi_1^3)+\frac{x^8}{12}[6+70\chi_1^2-3\chi_1^4+6\kappa_1(1+13\chi_1^2+6\chi_1^4
)+24(\chi_1+3\chi_1^3)B_2(\chi_1)]\bigg\},\nnm \\ &\text{where }\mathcal{F}_{\infty} = \frac{32}{5}\bigg(\frac{m_{2}}{M}\bigg)^2x^{10} \text{ and }\Omega \frac{dJ_1}{dt} = \frac{dm_1}{dt},
\end{alignat}
\end{widetext}
which is consistent with the results obtained via BH perturbation theory for the case of a tiny test particle orbiting a spinning BH in Refs.~\cite{Tagoshi:1997jy}. This result has been produced by solving the Teukolsky equation for a perturbation sourced by a tiny nonspinning BH ($M_{\mr{ext}}$) around a large spinning BH $M$ and computing the energy flux down the horizon. This result has also been reproduced in Ref.~\cite{Shah:2014tka} and the method has been employed to push the results for flux to infinity and horizon fluxes to a very high PN order in the test-body limit (see Refs.~\cite{Fujita:2011zk,Shah:2014tka,Fujita:2014eta}). It is thus reassuring that our expression for mass-loss matches with this in the appropriate limits, suggesting that our effective worldline picture is suitable for the purpose of modelling horizon-related dissipation in spinning BHs.
\section{Effect of horizon fluxes on the waveform for circular orbits}\label{sec:phase}
A simple way to derive the contribution to the waveform phase from horizon fluxes for parallel-spin--quasi-circular inspiral in the adiabatic limit is via the stationary-phase approximation (SPA)~\cite{Sathyaprakash:1991mt,Buonanno:2009zt,Arun:2008kb}. This was used to incorporate the effect of leading-order rate of change of mass (at 2.5PN) in the waveform in the appendix of Ref.~\cite{Brown:2007jx} and to 4 PN in Ref.~\cite{Datta:2020gem}, albeit the horizon fluxes used in Ref.~\cite{Datta:2020gem} (which were taken from Ref.~\cite{Alvi:2001mx}) are only accurate to leading order and the contribution due to the changing mass in the expression for binding energy was not taken into account [see Eq.~(\ref{eshift})]. Here, we use the same method (i.e., SPA) and extend the computation completely to 4PN order, while including all relevant effects due to the changing mass and spin and compute the phase contribution to the waveform due to horizon fluxes. We use $x$ as the gauge-invariant--PN-counting parameter to relative 1.5PN (and absolute 4PN) order. We start from the relation 
\begin{alignat}{3}
x =(M\omega_{\mr{orb}})^{\frac{1}{3}}= \bigg(M\frac{d\phi}{dt}\bigg)^{\frac{1}{3}},
\end{alignat}
where $\phi$ is the orbital phase and $M=m_1+m_2$ is the total mass of the system. The binding energy $E$ of the system is given to 1.5PN order\footnote{It is sufficient for us to include the expression for the binding energy to 1.5PN order as we are only interested in computing the waveform phase contribution due to horizon fluxes which were derived in this work to relative 1.5PN order.} (see, e.g., Ref.~\cite{Brown:2007jx}) as
\begin{alignat}{3}
\label{bend}\nnm E&=-\frac{M \eta x^2}{2}\bigg\{1+x^2\bigg(\frac{-3}{4}-\frac{\eta}{12}\bigg)\\& +x^3\bigg[\frac{8\delta \chi_a}{3}+\bigg(\frac{8}{3}-\frac{4\eta}{3}\bigg)\chi_s\bigg]\bigg\},
\end{alignat}
where $\delta = (m_1-m_2)/M$, $\eta=m_1m_2/M^2$, $\chi_a=(\chi_1-\chi_2)/2$, $\chi_s=(\chi_1+\chi_2)/2$.
Now, we use the energy balance law valid for circular orbits given by 
\begin{alignat}{3}
\dot{E}=-\mathcal{F}_{\infty}-\dot{M},
\end{alignat}
where $\mathcal{F}_{\infty}$ is the energy flux to infinity and over-dot represents derivative with respect to time. For noncircular orbits, one may have to include additional Schott terms~\cite{Bini:2012ji}. Now, we see from Eq.~(\ref{bend}) that $E$ is a function of $x$, the masses $m_1$, $m_2$ and the spins $\chi_1$, $\chi_2$, and thus we can write
\begin{alignat}{3}
\dot{E} &= \frac{\partial E}{\partial x} \dot{x} +  \frac{\partial E}{\partial m_1} \dot{m}_1  +  \frac{\partial E}{\partial m_2} \dot{m}_2   \nnm \\& +  \frac{\partial E}{\partial \chi_1} \dot{\chi}_1  +  \frac{\partial E}{\partial \chi_2} \dot{\chi}_2  ,
\label{eshift}
\end{alignat}
which, along with the balance relation, yields
\begin{alignat}{3}
\nnm \dot{x} &= -\bigg(\frac{\partial E}{\partial x}\bigg)^{-1}(\mathcal{F}_{\infty}+\dot{M}+\dot{m}_1\partial_{m_1} E + \dot{m}_2 \partial_{m_2} E  \\& +\dot{\chi}_1\partial_{\chi_1}E + \dot{\chi}_{2}\partial_{\chi_{2}E}),
\end{alignat}
where we can drop the terms arising from the spin-dependence as they do not contribute until relative 2.5PN order in horizon fluxes. Similarly, we can substitute the leading-order formula for $E$ in partial derivatives with respect to $m_1$ and $m_2$ and only substitute the flux to infinity up to 1.5PN relative order to get
\begin{alignat}{3}
\dot{x} = -\bigg(\frac{\partial E}{\partial x}\bigg)^{-1}[\mathcal{F}_{\mr{\infty}}^{\mr{1.5PN}}+\dot{M}-\frac{x^2}{2M^2}(m_{2}^2\dot{m}_1+m_1^2\dot{m}_{2})]. \nnm \\&
\end{alignat}
The flux to infinity up to 1.5PN order can  be found in Ref.~\cite{Brown:2007jx}. We can then invert this expression and integrate to compute the time function $t(x)$ and orbital phase $\phi(x)$ as
\begin{alignat}{3}
\delta t(x) = \int \frac{1}{\dot{x}} dx,\quad \int \frac{d\phi}{dt}dt =  \int \frac{x^3}{M} \frac{dt}{dx} dx =  \phi(x). \label{phiV}
\end{alignat}
Then, the orbital phase function $\phi(x)$ and the time function $t(x)$, can be related to the waveform phase $\psi$, in Fourier domain  for any spherical mode $m$, via the relation~\cite{Arun:2008kb}
\begin{alignat}{3}
\psi_{lm}(f)=2\pi f t_{f}- m \phi(t_f) - \frac{\pi}{4},
\end{alignat}
where $f$ is the Fourier variable (frequency) and $t_f$ corresponds to the time when the instantaneous GW frequency coincides with $f$, i.e.,
\begin{alignat}{3}
\frac{dm \phi}{dt} (t_f) = 2\pi f \implies x(t_f)=v=\bigg(\frac{2\pi M f}{m}\bigg)^{\frac{1}{3}}.
\end{alignat}
$\psi_{lm}(f)$ is a useful quantity directly relevant for detectors and we provide its correction due to the horizon fluxes explicitly as
\begin{alignat}{3}
\label{nfin}\nnm \delta\psi_{lm}(f) &= \frac{3}{128 \eta v^5}\frac{m}{2}\bigg[\sum_{n=5}^8 \delta\psi_{n}^{\mr{(PN)}}v^n \\& +\sum_{n=5}^8 \delta \psi_{n(l)}^{\mr{(PN)}}v^n  log(v)\bigg],  \\
v&=\bigg(\frac{2\pi M f}{m}\bigg)^{\frac{1}{3}},
\end{alignat}
where $\delta \psi_{lm}(f)$ is the correction to $\psi_{lm}(f)$ due to horizon fluxes with coefficients $\delta \psi_{n}^{(PN)}$ and $\delta \psi_{n(l)}^{(PN)}$ starting from 2.5PN ($n=5$) and up to 4PN ($n=8$) given by
\begin{widetext}
\begin{alignat}{3}
\delta \psi_5^{(\mr{PN})} &= -\frac{10}{9}[(1-3\eta)\chi_s(1+9\chi_a^2+3\chi_s^2)+\delta(1-\eta)\chi_a(1+3\chi_a^2+9\chi_s^2)],\\
\delta \psi_{5(l)}^{(PN)} &= 3 \delta \psi_5^{(\mr{PN})},\\
\delta \psi_7^{(\mr{PN})} &= \frac{5}{168}\{\delta\chi_a[-1667-4371\chi_a^2-13113\chi_s^2+616\eta^2(1+3\chi_a^2+9\chi_s^2)+5\eta(311+807\chi_a^2+2421\chi_s^2)]\nnm\\&+\chi _s \left[840 \eta ^2 \left(9 \chi _a^2+3 \chi _s^2+1\right)+\eta  \left(38331 \chi _a^2+12777 \chi
   _s^2+4889\right)-13113 \chi _a^2-4371 \chi _s^2-1667\right]\}, \\
   \delta \psi_{7(l)}^{(\mr{PN})} &= 0,
   \end{alignat}
   and 
   \begin{alignat}{3}
   \delta\psi_8^{(\mr{PN})}& = \delta\psi_8^{(\mr{PN}),a}+\delta\psi_8^{(\mr{PN}),b}+\delta\psi_8^{(\mr{PN}),c}\\
\delta \psi_{8(l)}^{(\mr{PN})} &= -3 \delta \psi_8^{(\mr{PN})},
\end{alignat}
with
\begin{alignat}{3}
\delta \psi_8^{(\mr{PN}),a} &= -\frac{5}{27}\{144 \pi  \delta  (\eta -1) \chi _a^3+48 \pi  \delta  (\eta -1) \chi _a+3 \left(278 \eta ^2-370 \eta +75\right)
   \chi _a^4+\left(-36 \eta ^2+213 \eta -67\right) \chi _a^2 \nnm \\ & +\chi _s \left[-12 \delta  \left(\eta ^2+190 \eta -75\right) \chi _a^3+2 \delta  \left(10 \eta ^2+124 \eta
   -67\right) \chi _a+432 \pi  (3 \eta -1) \chi _a^2+48 \pi  (3 \eta -1)\right] \nnm \\&+\chi _s^2 \left[432 \pi  \delta  (\eta -1) \chi _a+90 \left(36 \eta ^2-62 \eta +15\right) \chi _a^2-172 \eta
   ^2+303 \eta -67\right]+3 \left(82 \eta ^2-250 \eta +75\right) \chi _s^4\nnm \\&+\chi _s^3 \left[12 \delta  \left(21 \eta ^2-130 \eta +75\right) \chi _a+144 \pi  (3 \eta -1)\right]-12 \left(2 \eta ^2-4 \eta +1\right)\} ,
   \\ \delta \psi_8^{(\mr{PN}),b} &=-\frac{20}{9}\{[\delta(2\eta-1)\kappa_a+(-1-2\eta^2+4\eta)\kappa_s][1+6\chi_a^4+13\chi_s^2+6\chi_s^4+\chi_a^2(13+36\chi_s^2)]  \nnm \\& -2[\kappa_a(1-4\eta+2\eta^2)+\delta(1-2\eta) \kappa_s]\chi_a\chi_s [13+12(\chi_a^2+\chi_s^2)] \}
   \\ \delta \psi_8^{(\mr{PN}),c} &= \frac{80}{9}\{B_{2,s} [\left(2 \eta ^2-4 \eta +1\right) \chi _s \left(9 \chi _a^2+3 \chi _s^2+1\right)-\delta  (2 \eta -1) \chi _a
   \left(3 \chi _a^2+9 \chi _s^2+1\right)] \\&+B_{2,a}[3 \left(2 \eta ^2-4 \eta +1\right) \chi _a^3+9 \delta  (1-2 \eta ) \chi _a^2 \chi _s+\left(2 \eta ^2-4 \eta
   +1\right) \chi _a \left(9 \chi _s^2+1\right)-\delta  (2 \eta -1) \chi _s \left(3 \chi _s^2+1\right)] \}\nnm
\end{alignat}
\end{widetext}
where we have defined $\kappa_s=(\kappa_1+\kappa_2)/2$, $\kappa_a=(\kappa_1-\kappa_2)/2$, and $B_{2,s} = [B_2(\chi_1)+B_2(\chi_2)]/2$, $B_{2,a}= [B_2(\chi_1)-B_2(\chi_2)]/2$ for convenience. Note that the 2.5PN and 3.5PN corrections vanish for spinless case which is consistent with the fact that horizon fluxes only start at 4PN for nonspinning BHs (and 2.5PN for spinning case). We also find that the 4PN correction to the waveform phase contains functions that are nonpolynomial in the spin parameters through $\kappa = \sqrt{1-\chi^2}$ and $B_2 = \mr{Im}[\mr{PolyGamma}(0,3+i2\chi/\kappa)]$, as expected from the expression for the horizon energy fluxes (or rate of change of masses) in Eq.~(\ref{mlcircg}) at relative 1.5PN order. The Fourier phase solely due to the flux to infinity (i.e., neglecting horizon fluxes) to 3.5PN can be found in Eq.~(7) and Appendix~A in Ref.~\cite{Mehta:2022pcn}. The correction to the Fourier phase $\delta \psi_{lm}(f)$, can now be conveniently incorporated in waveform models to include the effect of horizon fluxes to next-to-next-to-leading order (up to 1.5PN relative, or 4PN absolute) during inspiral for quasi-circular aligned-spin binaries. 

\begin{table*}
\begin{center}
\begin{tabular}{ |c|c| } 
 \hline
 & $(10+10) M_{\odot}$, equal aligned spins $\chi_1=\chi_2=\chi$ \\ 
\hline \hline
 0PN & 603.6   \\ 
 1PN & 59.4  \\ 
 1.5PN & $-51.4 + 32\chi$ \\
 2PN & $4.1 - 4.4 \chi^2$  \\
 2.5PN & $-7.1 + 11.3 \chi + \boxed{10^{-3}12.8(1 + 3 \chi^2)\chi}$ \\
 3PN & $2.2 - 6.5 \chi - 0.64\chi^2 $\\ 
 3.5PN & $-0.8 + 3.6 \chi + 1.3 \chi^2 - 0.4 \chi^3 + \boxed{10^{-3}(8.4+22\chi^2)\chi}$ \\
 4PN & \boxed{10^{-3}[-0.3(1+\kappa)-(6.7+1.1 B_{2})\chi(1+3\chi^2)-(0.4+3.5\kappa)\chi^2+(9.5-1.6\kappa)\chi^4]} 	\\
 \hline
\end{tabular}
\label{equal}
\end{center}
 \caption{\label{equal} Number of orbital cycles as the frequency ($f$) of the gravitational wave increases from $f=10 \mr{Hz}$ to the frequency at the innermost stable circular orbit (ISCO) $f_{\mr{ISCO}}=1/(6^{3/2}M\pi) \mr{Hz}$ for equal masses (10 $M_{\odot}$) and equal aligned (to orbital angular momentum) spins. Recall that $\kappa=\sqrt{1-\chi^2}$. Here, $B_{2}=\mr{PolyGamma}(0,3+ 2i\chi/\kappa)$ lies between $0$ (at $\chi=0$) and $\pi/2$ (at $\chi=1$). The terms in boxes are contributions from horizon fluxes, and the remaining terms come from the flux to infinity, written here for comparison.}
\end{table*}
\begin{table*}
\begin{center}
\begin{tabular}{ |c|c| } 
 \hline
 & ($10$+$1.4$) $M_{\odot}$, (BH-NS), $\chi_1=\chi$, $\chi_2=0$ \\ 
\hline \hline
 0PN & 3587.6   \\ 
 1PN & 213.5  \\ 
 1.5PN & $-181.5 + 126.2\chi$ \\
 2PN & $9.8 - 13.5 \chi^2$  \\
 2.5PN & $-20.0 + 36.8 \chi + \boxed{10^{-2}9.4(1+3\chi^2)\chi}$ \\
 3PN & $2.3-18.6\chi-0.5\chi^2$\\ 
 3.5PN & $-1.8+10.5\chi+3.1\chi^2-\chi^3+ \boxed{10^{-2}(5.9+15.4\chi^2)\chi}$ \\
 4PN & \boxed{10^{-2}[-0.3(1+\kappa)-(4.3+1.2 B_{2})\chi(1+3\chi^2)-(1.7+3.9\kappa)\chi^2+(5.6-1.8\kappa)\chi^4]} 	\\
 \hline
\end{tabular}
\label{unequal}
\end{center}
 \caption{\label{unequal} Number of orbital cycles as the frequency increases from $f=10 \mr{Hz}$ to the frequency at the innermost stable circular orbit (ISCO) $f_{\mr{ISCO}}=1/(6^{3/2}M\pi) \mr{Hz}$ for a binary composed of a non-spinning 1.4 $M_{\odot}$ neutron star (NS) and a 10 $M_{\odot}$ BH with spin parameter $\chi$. Any contributions due to the internal structure of the NS have not been included.}
\end{table*}
To get a qualitative idea of the relevance of horizon fluxes to the waveform, we can look at the correction to the  orbital phase $\phi(x)$ and compute how many additional (or fewer) orbital cycles occur, as a result of the inclusion of those effects, for some specific choices of the initial masses of the BHs. We consider first the case of two initially equal-mass $m_1=m_2=10M_{\odot}$ and equal-aligned-spin $\chi_1=\chi_2=\chi$ BHs, and second the case for a binary consisting of a non spinning 1.4 $M_{\odot}$ neutron star and a 10 $M_{\odot} $ BH with spin parameter $\chi$. In the latter case, only the BH's horizon-flux contribution is considered and the neutron star is treated as a structureless particle. We then compute the additional (or fewer) number of cycles due to horizon flux(es) starting from the minimum lower frequency of the bandwidth of LVK detectors, $\omega=\pi\times10 \mr{Hz}$\footnote{Note that we are working in units where $G$=$c$=1.} to that of the innermost-stable circular orbit (ISCO) $\omega=\omega_{\mr{ISCO}} =1/(6^{\frac{3}{2}}M) \mr{Hz}$ (in Schwarzschild), which generally is a good approximation of the binary's merger frequency. We use the formula 
 \begin{alignat}{3}
   \mathcal{N}_{\mr{GW}}=\frac{\delta\phi[x=(M\omega_{\mr{ISCO}})^{\frac{1}{3}}]-\delta\phi[x=(M\pi\times10\mr{Hz})^{\frac{1}{3}}]}{\pi}, \label{number} \nnm \\&
 \end{alignat}
 where the correction to the orbital phase function due to horizon fluxes is obtained as shown in Eq.~(\ref{phiV}). We list the results obtained for the aforementioned special cases in Tables~\ref{equal} and \ref{unequal}. In Table~\ref{equal}, we also list the contribution to the number of cycles due to the flux to infinity up to 3.5PN (but with only nonspinning contributions to flux to infinity at 3PN and 3.5PN) using the expressions for fluxes and binding energy from Ref.~\cite{Brown:2007jx}. This is to facilitate comparison and get a qualitative understanding of the relevance of the horizon flux to the waveforms. Similar tables with flux-to-infinity contributions can be found for example in Refs.~\cite{Blanchet:2001ax, Blanchet:2006gy}. There are slight numerical differences between Table~\ref{equal} here and the tables in these works, because the final result is very sensitive to the precision used for the mass of the Sun and the gravitational constant. As clearly evident from the table, the contribution to the number of cycles from horizon fluxes (boxed terms in the table) is quite small when compared to the usual contributions from the flux to infinity even at the same PN order, although it is better for larger mass ratios. This is due to the (relatively) small numerical value of the coefficients in the horizon fluxes when compared with analogous terms in the fluxes to infinity when the masses are equal. This feature of small contribution due to the horizon fluxes has already been pointed out in Ref.~\cite{Alvi:2001mx} but it was obtained using an expression for the horizon fluxes that is only correct at leading order. This fact (relative smallness of horizon flux contributions) does not change much for other configurations of spins and masses either, at least for the frequency band of LVK detectors. Although the effect of the horizon flux on the waveform phase is small, they would need to be included when building highly accurate waveform models for next generation detectors on the ground and in space. 
\section{Conclusion}\label{sec:conclude}
In this work, we set out to tackle the problem of including
horizon-related dissipation effects in spinning BHs in an
effective worldline theory, which is an important physical effect to
include in precision GW predictions for future
detectors. For that purpose, we wrote down an effective action with
additional multipolar moment degrees of freedom which couple directly
with tidal fields in the action, and are tidally induced by them in
accordance with the symmetries of a Kerr BH, namely
axissymmetry and parity invariance. We fixed the remaining freedom in
the ansatz relating the tidal fields to the multipole moments by
considering a scattering scenario wherein GWs were
scattered off the effective particle and the degree of absorption was
compared with that obtained from the full theory by solving the
Teukolsky equation. A crucial ingredient in being able to fix the
complete dissipative part of the ansatz through this method was to
impose upon the effective theory the requirement that the scattering
be independent for different spheroidal modes of the Weyl scalar
$\psi_4$, which follows from the separability of the Teukolsky
equation in the full theory in spheroidal harmonics with spin weight
-2. Having fixed the relevant part (for horizon-related dissipation)
of the ansatz in this way, we used the model to compute the orbit
averaged variation in mass and spin due to horizon fluxes to relative
1.5PN order for a binary in circular orbit with parallel spins. The
mass and spin rate of change derived using our effective model is
consistent with the results obtained in the test-body limit in
Ref.~\cite{Tagoshi:1997jy} to relative 1.5PN order, and with
Ref.~\cite{Chatziioannou:2016kem} for generic mass ratios up to
relative 1PN order and at leading order with the generic mass ratio
results in Refs.~\cite{Goldberger:2020fot, Alvi:2001mx,
  Poisson:2005pi, Poisson:2004cw, Yunes:2005ve, Comeau:2009bz}. Importantly, we have weighed in one side (specifically on the side of Ref.~\cite{Tagoshi:1997jy}) in the previous discrepancy in the expression for evolution of mass in a binary in the literature between Refs.~\cite{Tagoshi:1997jy} and \cite{Chatziioannou:2016kem}. While the source of the earlier discrepancy is still unclear and remains to be settled, our approach suggests that it may have something to do with including the effect of octupolar tidal fields in the evolution of mass and spin.

Having consistently modelled the horizon-related dissipation and the associated changes in mass, spin and area of the horizon in this manner, we then proceeded to compute the contribution to the phasing of the waveform due to the relative 1.5PN horizon fluxes, which is relevant to the waveform at 4PN with respect to that of the leading-order quadrupolar flux to infinity. This was done using the SPA valid in the adiabatic quasi-circular regime of interest during inspiral. We found that a qualitative measure of the contribution of the horizon fluxes, i.e., the number of cycles in the waveform as the frequency evolves from $10\mr{Hz}$ to $f_{\mr{ISCO}}$
is very small (~2 to 3 orders of magnitude) compared to other contributions arising from GW energy flux to infinity at the same PN orders for typical masses observed by LIGO-Virgo-KAGRA detectors. 

An interesting future direction will be to use the model to derive the contribution to the waveform phasing without relying on the stationary-phase approximation to get a result valid outside of the adiabatic regime. This can be done for example by deriving the radiation-reaction forces due to the tidally-induced moments obtained from first-principles instead of relying on balance arguments. It is also of interest to consider how these results, namely the variation in mass and spin and the contribution to waveform phasing are affected in the presence of eccentricity or nonparallel spins. It may also be of interest to study possible resummations for the evolution equations of mass and spin and their contribution to the waveform phase along the lines of Refs.~\cite{Nagar:2011aa,Bernuzzi:2012ku,Taracchini:2013wfa}, now aided by an expression valid at higher (relative 1.5) PN orders for generic mass ratios. Finally, the approach used for modelling the particle in this work may be extended to generic compact bodies wherein the changes in mass and spin may occur due to tidal heating, e.g., in a viscous fluid. Parametrizing the changes in mass, spin and the subsequent contribution to waveform phasing for generic bodies could be very useful for testing the predictions of general relativity, and more specifically in the search for exotic compact bodies using next-generation GW detectors.
\acknowledgements
We thank Gustav Jakobsen, Chris Kavanagh, Gustav Mogull, Raj Patil and Khun Sang Phukon for useful discussions. We thank Chris Kavanagh in particular for sharing with us the knowledge for generating solutions to the Teukolsky equation.
\begin{widetext}
\appendix

\section{Scalar field scattering in effective worldline theory including leading-order tail effects}
\label{scalartail}
In the main text, we mentioned that the leading-order tail effect, due to the scattering of GWs off the particle's gravitational field leads to a factor of $(1+2\epsilon\pi)$ multiplying the leading-order degree of absorption. This is seen clearly in the Teukolsky solution given in Eq.~(\ref{tru2}), but not in the one derived using effective worldline theory in Eq.~(\ref{weftabs}) since we only solved the scattering problem in flat space. We motivated that this can be reproduced in the effective theory as well by including the effect of leading-order nonlinearities due to the gravitational field of the particle while solving the wave equation but did not prove it. Here, we show this explicitly in the case of a scalar field scattering off a spinless BH. In particular, we consider the scattering of the monopole mode $l=m=0$ to $\mathcal{O}(\epsilon^4)$ and show that an identical factor of $(1+2\epsilon\pi)$ multiplies the leading-order degree of absorption for this mode when the leading-order tail effects are included.

In the effective theory, we model the spinless BH as a particle with mass $m$ with an inducible monopole moment $m_{\phi}(\tau)$ in the presence of an external scalar field. We write an effective worldline action including a tidal scalar monopole moment as
\begin{alignat}{3}
S= -\int d\tau (m - K_{\mr{Q}}m_{\phi}(\tau) \phi) - \frac{K_{\phi}}{2}\int dt d^3 \vec{x} \sqrt{-g}g^{\alpha\beta}\nabla_{\alpha}\phi \nabla_{\beta}\phi + \frac{1}{16\pi G}\int d^4 x \sqrt{-g} R, 
\end{alignat}
which is identical to the action used in Ref.~\cite{Creci:2021rkz} except we have restricted to just including a monopole moment for simplicity.

Spherical symmetry ensures that the scalar monopole moment can only be induced by a scalar monopole mode. Since we are only interested in dissipative effects, we can write a general ansatz for the moment simply as
\begin{alignat}{3}
m_{\phi}(\tau) = GM K_\phi \sum_{n=0}^{\infty}(GM)^n \nu_n \frac{d^{2n+1}\phi}{d\tau^{2n+1}},
\end{alignat}

The particle sources a static gravitational field given at linear order in $G$ in the rest frame as 
\begin{alignat}{3}
h^{00} &= -4 \frac{GM}{r},~h^{0i} = h^{ij} = 0, \\
h^{\mu\nu} &	= \sqrt{-g}g^{\mu\nu} - \eta^{\mu\nu},
\end{alignat}
which will affect the behaviour of the scalar field through the Klein Gordon equation. The scalar field obeys the Klein Gordon equation in curved space-time, which to linear order in $G$ with the above metric perturbation is given by
\begin{alignat}{3}
\Box\phi&=-\ddot{\phi}+\nabla^2\phi = \frac{4GM}{r}\ddot{\phi} + \frac{K_Q}{K_{\phi}}m_{\phi}(\tau)\delta^{(3)}(\vec{r}),
 \\& \implies \omega^2 \phi +\nabla^2\phi = -\frac{4\epsilon\omega}{r}\phi + \frac{K_Q}{K_{\phi}}m_{\phi}(\tau)\delta^{(3)}(\vec{r}), 
\end{alignat}
in the rest frame of the particle. Here we have dropped divergent terms arising from the expansion of $(\sqrt{-g}-1)\times\delta^{(3)}(\vec{x})$ in the Klein-Gordon equation. Such terms can also be shown to cancel amongst themselves perturbatively but it is not relevant to our purpose. We have also restricted our attention to a single frequency mode of the wave, i.e., we set $\phi\sim\exp(-i\omega t)\psi(\vec{r})$. Now, we can expand the scalar field as 
\begin{alignat}{3}
\phi&=\phi^{(0)}+\epsilon \phi^{(1)}, \\
\phi^{(0)} &= C_{\mr{out}} \frac{\exp[-i\omega (t-r)]}{\omega r} + C_{\mr{in}}\frac{\exp[-i\omega (t+r)]}{\omega r} = C_{\mr{reg}}\frac{\sin(\omega r)}{\omega r} + C_{\mr{irr}} \frac{\cos(\omega r)}{\omega r}, 
\\ \Box\phi^{(0)}&= m_{\phi}(\tau)\delta^{(3)}(\vec{r}) 
\end{alignat}
where $\phi^{(0)}$ is the leading-order flat space-time solution and $\phi^{(1)}$ is the leading-order correction due to gravitational interaction. We have split $\phi^{(0)}$ into incoming ($C_{\mr{in}}$) and outgoing modes ( $C_{\mr{out}}$), and into regular ($C_{\mr{reg}}$) and irregular ($C_{\mr{irr}}$) modes. The regular mode is the homogeneous part of the flat space-time wave equation and the irregular part is the particular solution obtained from the source with the time-symmetric propagator. We can perturbatively write down an equation for $\phi^{(1)}$ as 
\begin{alignat}{3}
\Box \phi^{(1)} =  -\frac{4  \omega}{r}\phi^{(0)} +\mathcal{O}(\epsilon^2).
\end{alignat}
Solving this in general is difficult, but we only need to understand the asymptotic behaviour far away of $\phi^{(1)}$ and its behaviour at origin (location of the particle). This is because the asymptotic behaviour dictates the form of the wave as measured by a distant observer who can then measure the degree of absorption from that, and the value at origin perturbs the strength of the induced monopole moment through the ansatz. 

The general solution can be written as
\begin{alignat}{3}
\phi^{(1)} = \frac{\omega}{2\pi}\int d^3 \vec{r}'\bigg[ \frac{\phi^{(0)}(t-|\vec{r}'-\vec{r}
|,\vec{r}')}{r'|\vec{r}-\vec{r}'|}+ \frac{\phi^{(0)}(t+|\vec{r}'-\vec{r}
|,\vec{r}')}{r'|\vec{r}-\vec{r}'|}\bigg],
\end{alignat}
where we are using the time-symmetric propagator for consistency (as the irregular part of the leading-order solution contains both incoming and outgoing modes) and convenience. We can now derive its asymptotic behaviour as
\begin{alignat}{3}
\lim_{r\rightarrow\infty}\epsilon\phi^{(1)} &= \frac{ \epsilon\omega}{2\pi r}\int d^3 \vec{r}' \bigg[\frac{\phi^{(0)}(t-r+\hat{r}\cdot\vec{r}')}{r'} + \frac{\phi^{(0)}(t+r-\hat{r}\cdot\vec{r}')}{r'}\bigg], \\
&=4 \epsilon\exp(-i\omega t)\frac{\cos(\omega r)}{\omega r}\Big[C_{\mr{irr}}\int_0^{\infty} d\rho \frac{cos(\rho)\sin(\rho)}{\rho} + i C_{\mr{reg}}\int_0^{\infty} d\rho\frac{\sin^2(\rho)}{\rho}\Big], \\& = \exp(-i\omega t)\frac{\cos(\omega r)}{\omega r} \big(\pi \epsilon C_{\mr{irr}} + 4\epsilon i C_{\mr{reg}} \int_0^{\infty} d\rho \frac{sin^2(\rho)}{\rho}\big),
\end{alignat}
where the second integral next to $C_{\mr{reg}}$ which comes from the scattering of the homogenous solution off the static gravitational field of the particle is divergent but also does not contribute to absorption due to the $i$ in front of it. Thus, we can ignore it. The remaining part has the familiar $\pi\epsilon$ factor in front of it. Thus dropping the irrelevant part, we can write the total field asymptotically as
\begin{alignat}{3}
\lim_{r\rightarrow \infty} \phi =  \lim_{r\rightarrow \infty}[\phi^{(0)} + \epsilon \phi^{(1)}] = C_{\mr{reg}}\frac{\sin(\omega r)}{\omega r} + C_{\mr{irr}} (1+\epsilon \pi)\frac{\cos(\omega r)}{\omega r}.
\label{sasymp}
\end{alignat}
Now, before deriving the degree of absorption, we need to find the relation between $C_{\mr{reg}}$ and $C_{\mr{irr}}$ through the induced monopole moment. The strength of induced monopole moment depends on the value of the field at origin, and thus we also need to understand how the value of the field at origin is affected due to gravitational interaction. We will only use the regular part of the wave for computing the value of the field at origin since it is the input which induces the moment. Also, we are only interested in linear tidal effects in this work.

At origin, we have
\begin{alignat}{3}
\lim_{r\rightarrow 0} \epsilon \phi^{(1)} &= \frac{\epsilon \omega}{2\pi}\int d^3\vec{r}' \bigg[\frac{\phi^{(0)}_{\mr{reg}}(t-r',\vec{r}')}{(r')^2} + \frac{\phi^{(0)}_{\mr{reg}}(t+r',\vec{r}')}{(r')^2} \bigg] \\&= 4\epsilon \exp(-i\omega t)\bigg[C_{\mr{reg}} \int_0^{\infty} d\rho \frac{\cos(\rho)\sin(\rho)}{\rho} \bigg] \\&= \exp(-i\omega t)\pi \epsilon C_{\mr{reg}}.
\end{alignat}
Thus, we have for the total regular part of the field
\begin{alignat}{3}
\label{regch}
\lim_{r\rightarrow 0 } \phi^{(0)}_{\mr{reg}} = (1+\epsilon \pi)C_{\mr{reg}}\exp(-i\omega t),
\end{alignat}
which as claimed in the main text brings in another factor of $\epsilon\pi$. We can now compute the induced monopole moment as 
\begin{alignat}{3}
m_{\phi}(t) = -K_\phi GM\exp(-i\omega t)(1+\epsilon\pi)i C_{\mr{reg}}\sum_{n=0}^{\infty} \nu_{n} (-1)^n \epsilon^{2n+1} .
\end{alignat}
 Finally, we can now use the leading-order wave equation to solve for the relation between the regular and irregular coefficients as
\begin{alignat}{3}
\Box \phi^{(0)} &= \frac{4\pi}{\omega} \delta^{(3)}(\vec{r})C_{\mr{irr}} = \frac{K_Q}{K_{\phi}}m_{\phi}(t)\delta^{(3)}(\vec{r}), 
\\ \implies & C_{\mr{irr}} = \frac{K_Q}{K_{\phi}}m_{\phi}(t)\frac{\omega}{4\pi} = -(1+\epsilon\pi)\frac{K_Q}{4\pi}i C_{\mr{reg}} \epsilon^2
\sum_{n=0}^{\infty}\nu_n (-1)^n\epsilon^{2n} ,
\end{alignat}
which we can now substitute in Eq.~(\ref{sasymp}) to get
\begin{alignat}{3}
\lim_{r\rightarrow \infty} \phi &= \frac{\sin(\omega r)}{\omega r}C_{\mr{reg}} - \frac{\cos(\omega r)}{\omega r}(1+\epsilon\pi)^2\frac{K_Q}{4\pi}i C_{\mr{reg}} \epsilon^2
\sum_{n=0}^{\infty}\nu_n (-1)^n\epsilon^{2n}, \\& = \frac{\sin(\omega r)}{\omega r}C_{\mr{reg}} + \frac{\cos(\omega r)}{\omega r} C_{\mr{irr}}^{\mr{eff}}.
\end{alignat}
Note the factor of $(1+\epsilon\pi)^2 =  [1+2\epsilon\pi +\mathcal{O}(\epsilon^2)]$ modifying the effective values of the irregular part ($C_{\mr{irr}}^{\mr{eff}}$) of the wave far away from the source. This in turn modifies the coefficients next to the incoming and outgoing parts of the wave as well, changing the degree of absorption and the scattering phase.
The the degree of absorption can now be obtained as 
\begin{alignat}{3}
1-\bigg|\frac{C_{\mr{out}}^{\mr{eff}}}{C_{\mr{in}}^{\mr{eff}}}\bigg| = 1-\bigg|\frac{C_{\mr{reg}}+ i C_{\mr{irr}}^{\mr{eff}}}{C_{\mr{reg}}- i C_{\mr{irr}}^{\mr{eff}}}\bigg| = 1-\bigg|\frac{1+(1+2\epsilon\pi)\hat{K}_Q \epsilon^2 \sum_{n=0}^{\infty}\nu_n(i\epsilon)^{2n}}{1-(1-2\epsilon\pi)\hat{K}_Q \epsilon^2 \sum_{n=0}^{\infty}\nu_n(i\epsilon)^{2n}}\bigg|,
\end{alignat}
where $C_{\mr{in/out}}^{\mr{eff}}$  are the coefficients next to the incoming/outgoing parts of the complete solution at asymptotic infinity. We have also defined $\hat{K}_Q = K_Q/(4\pi)$. Note that in the absence of tail corrections, there are no odd powers of $\epsilon$ in the degree of absorption for a spinless BH. Now, expanding this in $\epsilon$, we get
\begin{alignat}{3}
1-\bigg|\frac{C_{\mr{out}}^{\mr{eff}}}{C_{\mr{in}}^{\mr{eff}}}\bigg| = -2 \hat{K}_Q \epsilon^2(1+2 \epsilon \pi) + \mathcal{O}(\epsilon^4),
\end{alignat}
where we have truncated our expression to next-to-leading order since we only included leading-order tail effects in this analysis. Here, we see explicitly that the leading-order tail effect, arising from the scattering of the wave off the static gravitational field of the particle modifies the leading-order degree of absorption by a factor of $(1+2\epsilon\pi)$. Crucially, this introduces odd powers of $\epsilon$ as well which the effective theory cannot otherwise reproduce. We have checked it against the same result obtained in the real theory by solving the Klein Gordon equation in the vicinity of a real BH with incoming boundary conditions at the horizon and obtained
\begin{alignat}{3}
8\epsilon^2(1+2\pi\epsilon) + \mathcal{O}(\epsilon^4),
\end{alignat}
as the degree of absorption for the monopole mode of a scalar wave. Note that this has a form identical to that obtained from the effective theory when leading-order tail effects are included, thus proving our claim in the main text for the special case of monopolar scalar-field scattering.
\label{regcorr}

\end{widetext}
\end{document}